\documentclass[jacsat,article,12pt,superscriptaddress]{revtex4-2} %aps,showpacs,twocolumn
\usepackage{amsmath,amsfonts,bm}
\usepackage[dvips]{graphicx}
\usepackage[all]{xy}
\usepackage{color}
\usepackage{color,soul}
\usepackage{diagbox}
\usepackage{amssymb}
\usepackage{physics}
\def\:{\colon\!}
\usepackage [autostyle, english = american]{csquotes}
\MakeOuterQuote{"}

\begin{document}

\title{Pump-intensity-scaling of Two-Photon-Absorption and Photon Statistics of Entangled-Photon Fields
}

\author{Deependra Jadoun}
\affiliation{Department of Chemistry, University of California, Irvine, CA 92614, USA}
\affiliation{Department of Physics and Astronomy, University of California, Irvine, CA 92614, USA}
\author{Upendra Harbola}
\affiliation{Department of Inorganic and Physical Chemistry, Indian Institute of Science, Bangalore 560012,  India}
\author{Vladimir Y. Chernyak}
\affiliation{Department of Chemistry, Wayne State University, 5101 Cass Ave, Detroit, Michigan 48202, USA}
\affiliation{Department of Mathematics, Wayne State University, 656 W. Kirby, Detroit, Michigan 48202, USA}
\author{Shaul Mukamel}
\affiliation{Department of Chemistry, University of California, Irvine, CA 92614, USA}
\affiliation{Department of Physics and Astronomy, University of California, Irvine, CA 92614, USA}

\date{\today}

\begin{abstract}
We use a non-perturbative theoretical approach to the parametric down-conversion (PDC) process,
which generates entangled-photon field for an arbitrarily strong pump-pulse.
This approach can be used to evaluate multi-point field correlation functions
to compute nonlinear spectroscopic signals induced by a strong pump.
The entangled-photon statistics is studied using Glauber's $g^{(2)}$ function,
which helps understand the significance of the photon
entanglement-time and the pump-pulse intensity on spectroscopic signals.
Under the non-perturbative treatment of the entangled field,
the two-photon absorption (TPA) signal shows linear to strongly non-linear growth with the pump intensity,
rather than linear to quadratic scaling reported previously.
An increase
in the range of pump intensity for the linear scaling is observed as the pump band-width is increased. We propose an experimental scheme that can select
contributions to the TPA signal that arise solely from interactions with the
entangled photons, and filter out unentangled photon contributions,
which are dominant at higher pump intensities, paving a way to explore the entanglement effects at higher intensities.

\end{abstract}


\maketitle
\section{Introduction}
Entanglement is a quantum mechanical effect that correlates two
or more particles in a non-classical way. An entangled pair of photons, also known as quantum light, is one such example where the time-energy entanglement between the two entangled photons can be utilized in spectroscopy
to gain insights into the chemical dynamics of molecules at an uprecedented resolution that is
not possible with classical light \cite{roadmap}.
Two-photon absorption (TPA) is the simplest spectroscopic technique that can demonstrate the merits of quantum light \cite{GillesPRA1993,PRL1989,OtEx2021}. 
The entangled-photon pairs obtained by parametric-down 
conversion (PDC) have been employed in numerous 
TPA experiments \cite{JimensasePRApp2021,LeeJPCB2006, GoodsonJPCL2013, GoodsonJPCL2017, GoodsonJACS2009,hickam22jpcl,corona22jpca,tabakaev21pra}.  
The signal is commonly detected by the fluorescence from doubly excited molecular states \cite{GuzmanJACS2010, LeeJPCB2006,TabakaevPRL2022,LandesPRR2021}.

Most theoretical studies involving entangled-photons
are limited to the weak pump field regime, which involves
a single pair of entangled-photons interacting with the molecule
\cite{MatthiasPNAS2023,MonsalveJPCA2017,BanaclochePRL1989}.
It is experimentally challenging to observe the weak spectroscopic signal
generated by isolated entangled-photon pairs\cite{LandesArXive2024, RaymerJCP2021,MikhaylovJPCL2022}.
This difficulty may be overcome by increasing the pump intensity, whereby several
entangled-photon pairs can interact with the molecule.
One such example of the entangled-field is the bright-squeezed vacuum \cite{Raymer2022}. 
However, a strong pump
increases unwanted contributions that arise from molecular interactions with
photons belonging to different entangled pairs that lack the 
quantum information of the entangled field. 

A spectroscopic signal generated by the bright squeezed vacuum can be
divided into two parts, denoted as inter-mode and intra-mode
contributions, where the former involves both the
"signal" and the "idler" modes \cite{svozilik18cp}, while the latter has only the "signal" or the "idler" mode contribution. Only the inter-mode part is sensitive to the photon entanglement. 
We have recently shown that at low pump intensities (but beyond single entangled-pair limit), the light-to-matter entanglement transfer is reduced by the presence of intra-mode contributions \cite{UH-AvsQuantum2023}.
Recently, intense entangled beams have been employed in TPA
\cite{svozilik18cp} and
virtual-state spectroscopy\cite{svozilik18josab} to study the impact 
of the time-entanglement between the two modes.
The intra-mode contributions dominate at large pump intensities, which can obscure
the quantum effects arising from the entangled nature of the field.
This causes difficulties in observing the theoretically predicted
effects of photon-entanglement in experiments.
An accurate description of the entangled field effects,
therefore, requires using arbitrary pump-pulse intensity (to enhance the signal) together with a spectroscopic scheme that can remove the unwanted background intra-mode contributions. This is the aim of the present work.

We use a non-perturbative approach to describe the pump field with arbitrary spectral width and intensity, which interacts with the PDC crystal to generate entangled-photon field. 
Starting with the effective Hamiltonian previously used by Dayan and Raymer 
\cite{DayanPRA2007,Raymer2022} to study high pump-intensity effects, we derive a system of
linear integro-differential equations whose solutions with appropriate boundary conditions provide the exact correlation functions of the
entangled field. Our system of equations is equivalent to the one presented in Ref. \cite{SilberhornNJP2013}.
Switching to the Wigner representation of the field correlation functions allows us to extend the exact analytical solutions for
infinitely narrow spectral width
pump (presented, e.g., in Ref. \cite{Raymer2022}) to obtain an asymptotically exact 
expression for the case of a narrow, rather than an infinitely narrow, pump. We compare the “almost analytical” Wigner approach with the results obtained
from a completely numerical solution 
and (i) show that for a sufficiently narrow pump, the two methods show excellent 
agreement, and (ii) identify the range of the pump spectral width, 
where the narrow pump solution can be safely implemented.

We use this approach to
evaluate the Glauber's $g^{(2)}$ function of entangled photons,
and study the role of entanglement-time and pump spectral width
on the photon-statistics of the entangled field. 
We further investigate the TPA signal and its
scaling with the pump intensity and spectral width. We finally propose an experimental 
protocol that can remove the unwanted intra-mode contributions from the TPA signal.

\section{Theory}
The TPA signal is given by the photon creation rate in the output (fluorescence)
mode with frequency $\omega_f$, (see Sec. I in the SI),
\begin{equation}
    \begin{aligned}
        R(\omega_f,t) & \equiv \sum_{\lambda}\frac{d}{dt}\langle E_{f}^{\lambda\dag}(t)E_{f}^\lambda(t)\rangle \\ &
 = \frac{2}{\hbar^2}\Re\sum_{\lambda}\sum_{i>j}\sum_{i^\prime > j^\prime} (\mu_{ij}\cdot \epsilon_\lambda) 
 (\mu^{*}_{i^\prime j^\prime}\cdot \epsilon_\lambda)
\int_{-\infty}^t d\tau \langle B_{ijR}^\dag(t) B_{i^\prime j^\prime L}(\tau)\rangle e^{-i\omega_f(t-\tau)}
    \end{aligned}\label{eq-1}
\end{equation}
Here $\mu_{ij}$ is the 
transition dipole moment between molecular states $i$ and $j$, and $\epsilon_\lambda$ is the output mode polarization. 
$B_{ij}=|j\rangle\langle i|$ is the excitation operator from the $i$th to $j$th electronic molecular state, and the subscripts $R$ and $L$ denote "right" and "left" superoperators, respectively, defined in Liouville space through their actions on the density-matrix, [$e.g$, $A_L\rho = A\rho, A_R\rho = \rho A$ such that $A_-\rho=(A_L-A_R)\rho$]\cite{NEGF08}. The angular bracket $\langle\cdot\rangle$
denotes the trace over the combined molecule and the incoming entangled field degrees-of-freedom.

Computing $R(\omega_f,t)$ requires calculating the exciton correlation function
$\langle B_{ij R}^\dag(t) B_{i^\prime j\prime L}(\tau)\rangle$. 
This can be done by using a modified correlation 
${\cal G}_{iji^\prime j^\prime}^{RL}(t,\tau)=\langle T B_{ijR}^\dag(t) B_{i^\prime j^\prime L}(\tau)\rangle$,
where $T$ is the time-ordering operator. $R(\omega_f)$ depends on the retarded ($t>\tau$) correlation function.  
Calculating  ${\cal G}^{RL}_{iji^\prime j^\prime}(t,\tau)$ requires computationally expensive
solution of several self-consistent equations in terms of single-particle (Green's functions) propagators\cite{Bethe-Salpeter,NEGF06, NEGF08} for the molecule. 
In the following, we compute the exciton correlation function perturbatively in the (incoming) entangled field-molecule coupling by expressing the correlation function in the interaction picture.
\begin{equation}
{\cal G}_{iji^\prime j^\prime}^{RL}(t,\tau) = 
\langle T B_{ijR}^\dag(t) B_{i^\prime j^\prime L}(\tau) e^{-\frac{i}{\hbar}\int d\tau_1H_{int-}(\tau_1)}\rangle\label{eq-0}
\end{equation}
with $H_{int}(t)=\sum_{ijq\lambda}\mu_{ij}B_{ij}^\dag(t)E_q^\lambda(t)+h.c.$, where $E_q^\lambda$ is the annihilation operator for the 
 the incoming field with frequency $\omega_q$ and polarization $\epsilon_\lambda$. All time-dependences are in the interaction picture. 

\begin{figure}[h]
\includegraphics[scale=0.32]{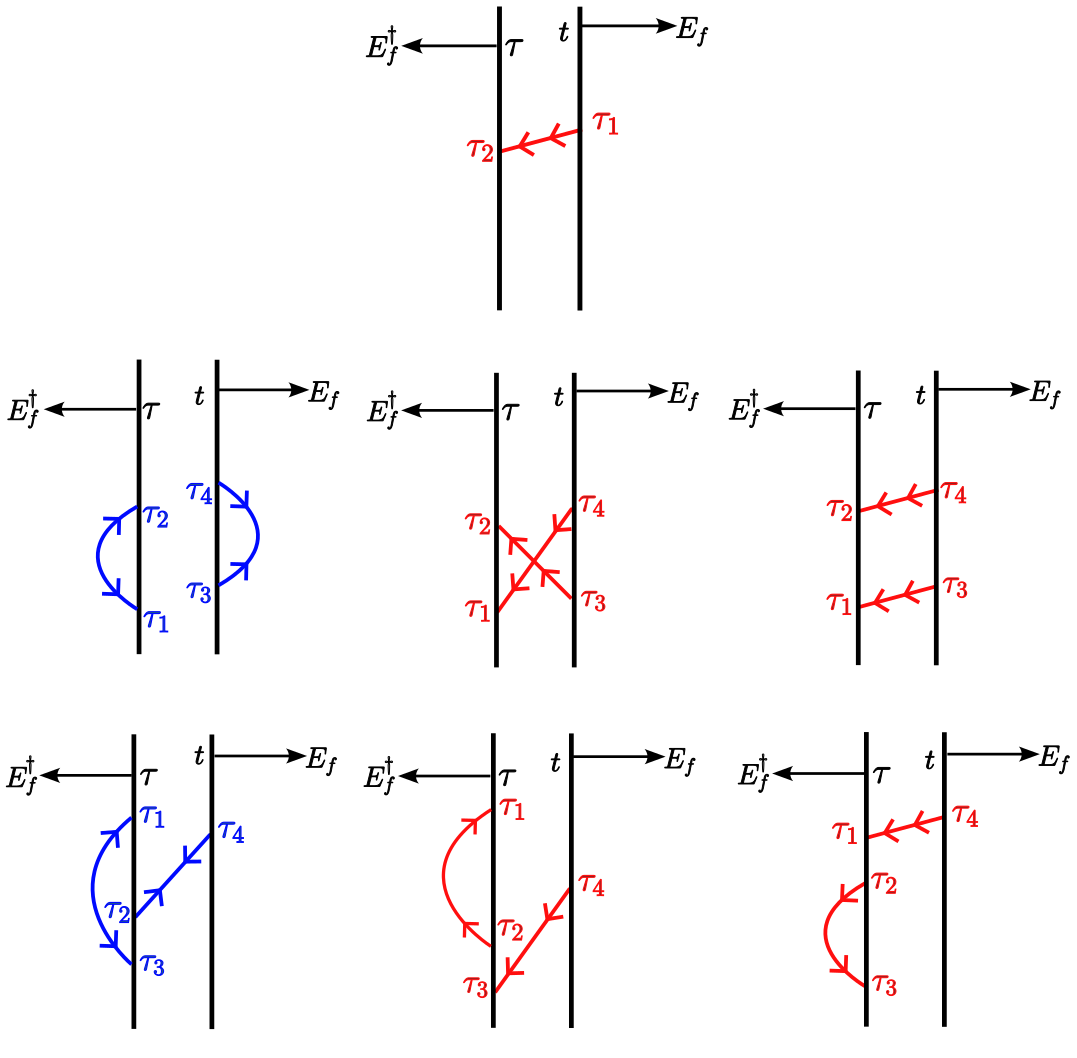}
\caption{Diagrams contributing to the first term (top panel), the second term (middle panel), and the third term (bottom panel) in Eq. (\ref{eq-1a}). Diagrams for the last term in Eq. (\ref{eq-1a}) are obtained by interchanging the left and right interactions at $\tau_i, i=1,2,3,4$, in the bottom diagram. Interactions at times $t$ and $\tau$ represented by the horizontal arrows correspond to the observed fluorescent mode. Red (blue) curves with two arrowheads denote ordinary (anomalous), $D^{-+} (D^{--}, D^{++})$ entangled-field propagators (see discussion below Eq. (\ref{eq-1b})). Arrowheads pointing (out) into the vertical black lines represent field $(E^\dag) E$.} 
\label{fig-a}
\end{figure}

The zeroth-order contribution in the perturbative expansion of Eq. (\ref{eq-0})
in $H_{int}$ vanishes for a molecule initially in the ground state. The two 
leading (second and fourth) -order contributions are, 
\begin{widetext}
\begin{eqnarray}
\label{eq-1a}
&&{\cal G}_{iji^\prime j^\prime}^{RL}(t,\tau) \approx 
\frac{2}{\hbar^2}(\mu_{i_1j_1}\cdot\epsilon_{\lambda_1})(\mu_{i_2j_2}^{*}\cdot\epsilon_{\lambda_2}) 
\int d\tau_1d\tau_2 \left[\frac{}{}\right.\nonumber\\
&&\left.\langle T E_{q_1L}^{\lambda_1}(\tau_1)E_{q_2R}^{\lambda_2\dag}(\tau_2)\rangle 
\langle T B_{ijR}^\dag(t) B_{i^\prime j^\prime L}(\tau) B^\dag_{i_1j_1L}(\tau_1)B_{i_2j_2R}(\tau_2)\rangle\right.\nonumber\\
&+& \left.\frac{3}{\hbar^2}
(\mu_{i_3j_3}^*\cdot\epsilon_{\lambda_3})(\mu_{i_4j_4}^{*}\cdot\epsilon_{\lambda_4})
 \int d\tau_3 d\tau_4  
\langle T E_{q_1L}^{\lambda_1}(\tau_1)E_{q_2L}^{\lambda_2}(\tau_2)E_{q_3R}^{\lambda_3\dag}(\tau_3)E_{q_4R}^{\lambda_4\dag}(\tau_4)\rangle \right.\nonumber\\
&\times& \left.\langle T B_{ijR}^\dag(t) B_{i^\prime j^\prime L}(\tau) B^\dag_{i_1j_1L}(\tau_1)B^\dag_{i_2j_2L}(\tau_2)B_{i_3j_3R}(\tau_3)
B_{i_4j_4R}(\tau_4)\rangle
\right.\nonumber\\
&-& \left.
\frac{6}{\hbar^2}
(\mu_{i_3j_3}\cdot\epsilon_{\lambda_3})(\mu_{i_4j_4}^{*}\cdot\epsilon_{\lambda_4})
 \int d\tau_3 d\tau_4 \right.\nonumber\\
  &\times&\left.\left(\frac{}{}
\langle T E_{q_1L}^{\lambda_1}(\tau_1)E_{q_2L}^{\lambda_2}(\tau_2)E_{q_3L}^{\lambda_3\dag}(\tau_3)E_{q_4R}^{\lambda_4\dag}(\tau_4)\rangle
 \langle T B_{ijR}^\dag(t) B_{i^\prime j^\prime L}(\tau) B^\dag_{i_1j_1L}(\tau_1)B_{i_2j_2L}(\tau_2)B_{i_3j_3L}^\dag(\tau_3)B_{i_4j_4R}(\tau_4)\rangle
 \right. \right.\nonumber\\
 &+&\left.\left.
\langle T E_{q_1L}^{\lambda_1}(\tau_1)E_{q_2R}^{\lambda_2}(\tau_2)E_{q_3R}^{\lambda_3\dag}(\tau_3)E_{q_4R}^{\lambda_4\dag}(\tau_4)\rangle
 \langle T B_{ijR}^\dag(t) B_{i^\prime j^\prime L}(\tau) B^\dag_{i_1j_1L}(\tau_1)B_{i_2j_2R}(\tau_2)B_{i_3j_3R}^\dag(\tau_3)B_{i_4j_4R}(\tau_4)\rangle 
 \right)\right]\nonumber\\
\end{eqnarray}
\end{widetext}
where summation over repeated indices is implied. 
The physical processes which give rise to the various
contributions in Eq. (\ref{eq-1a}) are depicted in Fig. (\ref{fig-a}).

The first term represents an excitation from the ground state $|g\rangle$ to the doubly excited state $|f\rangle$ by the absorption of a 
resonant pump photon. It could also represent the absorption of a single "idler" or 
"signal" photon, which has the proper energy due to the finite pump-pulse bandwidth. 
Hereafter, we assume that the pump-photons in the
crystal output are filtered out using phase-matching and do not 
contribute to the first term.
Note that each field operator includes both the "signal" and "idler" modes, 
$E_{q}^\lambda=E_s^\lambda(\omega_q)+E_i^\lambda(\omega_q)$.
The field correlation function  $\langle T E_{q_1L}^{\lambda_1}
(\tau_1)E_{q_2R}^{\lambda_2\dag}(\tau_2)\rangle$ (Section III in the SI)
has four terms, where two cross terms that contain inter-mode correlations vanish. 
Thus, the first term in Eq. (\ref{eq-1a}) only includes intra-mode photon 
correlations which is non-zero if the polarizations ($\epsilon_{\lambda_1}$
and $\epsilon_{\lambda_2}$) are parallel.

The second and the third terms in Eq. (\ref{eq-1a}) contain four-time field correlation functions that, by use of Wick's theorem for Boson fields, may be 
factorized into products of two-time field correlation functions.
\begin{equation}
    \begin{aligned}
        D_{\lambda_1\lambda_2\lambda_3\lambda_4}^{--++LLLR}(\tau_1,\tau_2,\tau_3,\tau_4) &=
\langle T E_{q_1L}^{\lambda_1}(\tau_1)E_{q_2L}^{\lambda_2}(\tau_2)E_{q_3L}^{\lambda_3\dag}(\tau_3)E_{q_4R}^{\lambda_4\dag}(\tau_4)\rangle \\ &
=\langle T E_{q_1L}^{\lambda_1}(\tau_1)E_{q_2L}^{\lambda_2}(\tau_2)\rangle\langle TE_{q_3L}^{\lambda_3\dag}(\tau_3)E_{q_4R}^{\lambda_4\dag}
(\tau_4)\rangle \\ & + \langle T E_{q_1L}^{\lambda_1}(\tau_1)E_{q_3L}^{\lambda_3\dag}(\tau_3)\rangle\langle TE_{q_2L}^{\lambda_2}(\tau_2) E_{q_4R}^{\lambda_4\dag}
(\tau_4)\rangle \\ & + \langle T E_{q_1L}^{\lambda_1}(\tau_1)E_{q_4R}^{\lambda_4\dag} (\tau_4) \rangle\langle TE_{q_2L}^{\lambda_2}(\tau_2) E_{q_3L}^{\lambda_3\dag}
(\tau_3)\rangle
    \end{aligned}\label{eq-1b}
\end{equation}
where a "- (+)" sign on the propagator $D$ denotes that the corresponding field operator is $E (E^\dag)$, for example, 
$D^{-+}(t,t^\prime)=\langle T E(t)E^\dag(t^\prime)\rangle$. 
The first term in Eq. (\ref{eq-1b}) is non-zero only when the paired modes, $(q_1,q_2)$ and $(q_3,q_4)$, are different "signal" and "idler" modes.
This contribution, therefore, depends on inter-mode correlations and carries 
information on the quantum state of both modes. 
The other two terms survive only when both 
paired modes belong to the same mode and represent intra-mode field correlations. 

\section{Model simulations}

\begin{figure}[h]
    \centering
    \includegraphics[width=0.3\linewidth]{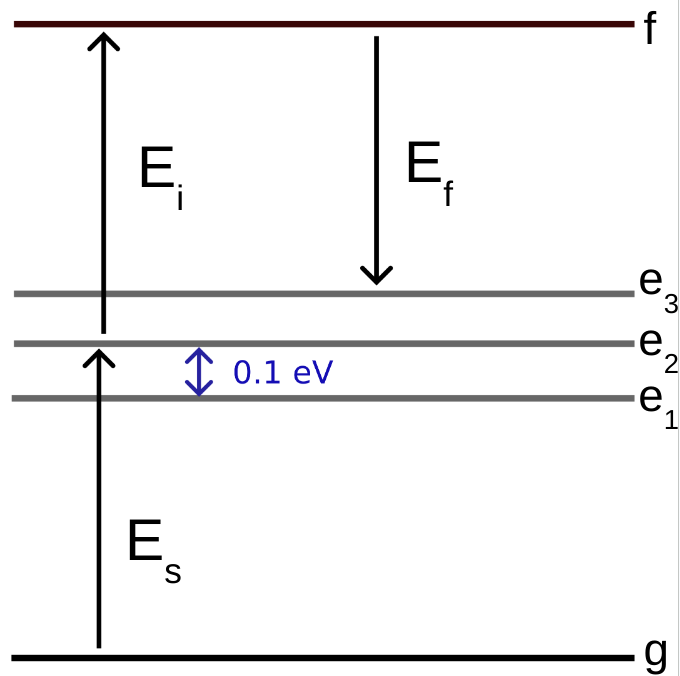}
    \caption{Energy level scheme used in our simulations with energies
    $e_1$=1.9\,fs$^{-1}$, $e_2$=2\,fs$^{-1}$, $e_3$=2.1\,fs$^{-1}$, and $e_4$=4\,fs$^{-1}$.} 
    \label{fig:ls}
\end{figure}

We consider the molecular level-scheme depicted in Fig. (\ref{fig:ls}),
which consists of a ground state $|g\rangle$,
one doubly excited state $|f\rangle$, and three intermediate singly 
excited states $|e_i\rangle, i=1,2,3$. 
The signals arising from 
the various processes depicted in Fig. (\ref{fig-a}) are computed
by expanding the molecular correlation functions in Eq. (\ref{eq-1a}) in
molecular eigenstates, as discussed in  Sec. II in the SI. 
The leading-order term, given by the first diagram 
in Fig. (\ref{fig-a}), originates from intra-mode interactions. 
The other terms contain both the inter- and the intra-mode  
contributions, represented by the blue and the red colored field propagators, 
respectively. These field propagators are computed using the
effective Hamiltonian approach given in Sec. III in the SI. The different  
contributions, $R^{(n)}$, given in Eqs. (5-10) in SI, therefore, carry 
quantum information regarding the entangled light field and are
related to Glauber's $g^{(2)}$ function which is commonly used to describe the non-
classical behavior of light\cite{Glauber}.  $g^{(2)}(t,\tau)$ can be measured 
by the coincidence-counting of photons at times $t-\tau$ and $t$,
 \begin{eqnarray}
 \label{g2}
 g^{(2)}(t,\tau) =\frac{\langle E^\dag(t-\tau)E^\dag(t)E(t)E(t-\tau)\rangle}{\langle E^\dag(t-\tau)E(t-\tau)\rangle \langle E^\dag(t)E(t)\rangle}.
 \end{eqnarray}
 Integrating over $t$ gives,
 \begin{equation}
 \label{g2-1}
 g^{(2)}(\tau) = \frac{1}{S(\tau)} \int \int \int \frac{d\omega_1d\omega_2 d\omega_3}{(2\pi)^3}   e^{-i(\omega_2-\omega_3)\tau} 
\langle E^\dag(\omega_1)E^\dag(\omega_2)E(\omega_3)E(\omega_1+\omega_2-\omega_3)\rangle
 \end{equation}
  where $\omega_4=\omega_1+\omega_2-\omega_3$ and $S(\tau)=\int \frac{d\omega_1d\omega_2d\omega_3}{(2\pi)^3} \langle E^\dag(\omega_1)E(\omega_4)\rangle \langle E^\dag(\omega_2)E(\omega_3)\rangle e^{-i(\omega_2-\omega_3)\tau}$ is a normalization factor that only includes 
  the intra-mode field correlations.   
$g^{(2)}$ can be decomposed into a sum of intra-mode, $g_0^{(2)}$, and inter-mode, $g_1^{(2)}$, contributions. Both contributions decay with $\tau$. 
At long $\tau$, the normalized intra-mode part saturates to unity,
while the inter-mode part vanishes. 

  %%%%%%%%%%%
\begin{figure}[h]
\includegraphics[scale=0.5]{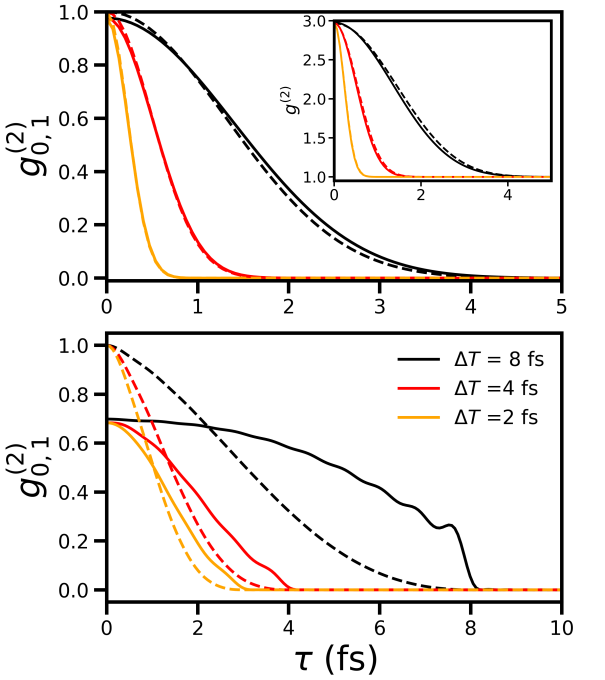}
\caption{Time-dependence of the inter-mode ($g^{(2)}_1$, solid) and intra-mode ($g^{(2)}_0-1$, dashed) parts of $g^{(2)}(\tau)$ for different entanglement times, as indicated. Upper(lower) panel: for high (low) pump intensity, $I_p$=3.35$\times$10$^{16}$ (3.35$\times$10$^{14}$)\,W/cm$^2$. Other parameters are: $\sigma_p$=0.3\,fs$^{-1}$, $\omega_p$=4\,fs$^{-1}$, $\omega_s$=2\,fs$^{-1}$, and PDC crystal length $l=20$\,$\mu$m. The inset compares the $g^{(2)}(\tau)$
function calculated numerically (solid) with the semi-analytical (dashed) solution.}
\label{fig-g2}
\end{figure}
\begin{figure}[h]
\includegraphics[scale=0.5]{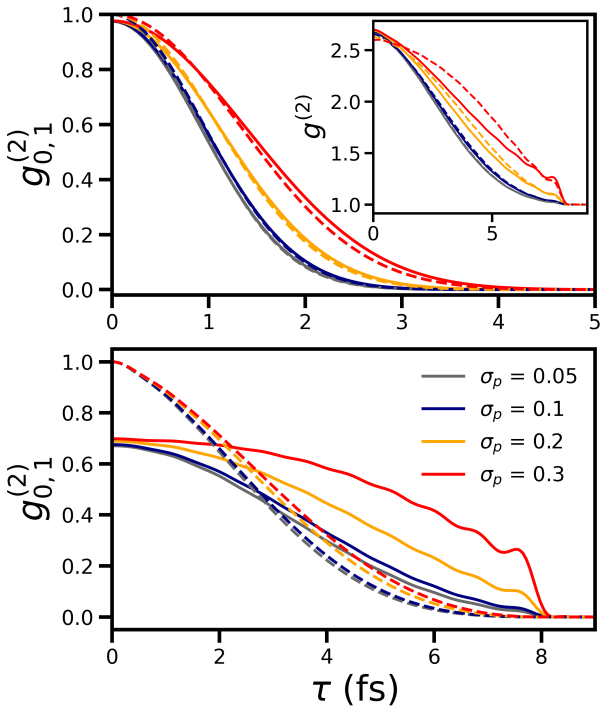}
\caption{
The time-dependence of the inter-mode ($g^{(2)}_1$, solid) and the intra-mode ($g^{(2)}_0$, dashed) contributions 
for different pulse bandwidths $\sigma_p$. The upper (lower) panel 
shows the inter-mode and intra-mode contributions
for a strong (weak) pump with the field strength of $I_p$=3.35$\times$10$^{16}$ (3.35$\times$10$^{14}$)\,W/cm$^2$ for the entanglement time $\Delta T=8$fs. The inset compares the total $g^{(2)}$ calculated numerically
(solid) with semi-analytical results (dashed) for $I_p$=3.35$\times$10$^{14}$\,W/cm$^{2}$.}
\label{fig-g2-2}
\end{figure}
 %%%
 
The inter- and intra-mode components of 
$g^{(2)}(\tau)$ for various field entanglement times are 
displayed in Fig. (\ref{fig-g2}). Both components show strong dependence on the 
entanglement time and rapidly decay
as the entanglement time is decreased. At high pump intensities, the two 
contributions are virtually identical. However, differences show up at low pump 
intensities where the inter-mode contribution survives for longer delays and vanishes 
beyond the entanglement time. For a degenerate entangled field produced by a 
continuous laser field, the Glauber function is computed in Sec. III-B in the SI. 
The temporal profile of $g^{(2)}_{0/1}(\tau)$
is Gaussian with a time-scale, which is mainly determined
by the entanglement time for low-pump intensities, while, at larger
intensities, it decreases with the
intensity, leading to a shorter effective entanglement time. The analytic
results for the continuous field can be extended for a finite (non-zero) but large 
time-scale (narrow band-width) pump pulse using Wigner representation as discussed in Sec. III in the SI.
The inset in
Fig. (\ref{fig-g2}) compares the numerical results with the
semi-analytic (Wigner) results for $\sigma_p=0.3$ fs$^{-1}$.

The time-dependent Glauber function is depicted for different pump bandwidths
in Fig. (\ref{fig-g2-2}). At high pump intensities, as the bandwidth is increased, the 
relaxation slows down, implying a longer effective entanglement time.
The difference between the intra- and inter-mode components is more 
pronounced at low pump intensities.
The relaxation of the inter-mode contribution strongly depends on the 
bandwidth, which is qualitatively different from the intense pump case.
As the bandwidth
is increased, it shows weaker dependence on $\tau$ for $\tau$ smaller than the 
entanglement time, and vanishes for larger $\tau$. The relaxation in the 
intra-mode part also slows down with increased bandwidth but remains 
qualitatively the same as for the intense pump. Thus, the inter-mode part of the 
Glauber function directly reveals information on the field entanglement time 
at larger bandwidths and low pump intensities. The inset compares
the numerical and the semi-analytic results for a finite band-width pump.
The exact numerical method and the semi-analytic Wigner approach agree well for small $\sigma_p \Delta T < 0.2$.

 \subsection{The TPA and its intensity scaling for a narrow-band pump} 

Having discussed the statistical properties of the entangled field, we now turn
to the TPA process, detected by fluorescence.
All processes depicted in Fig. (\ref{fig-a}) generally contribute to 
the signal. To simplify the analysis, we focus on a narrow-band pump 
resonant with the doubly excited state of the molecule. 
The "signal" and "idler" photon energies are bounded by the double-excitation energy.
The doubly-excited state population is generated by absorbing the "signal" and
 "idler" photons and the fluorescence is given by the $R^{(2)}$ term 
in Eq. 6 of the SI, the correspondsing diagrams are given
in the second row in Fig. (\ref{fig-a}). 
The radiative transitions 
 $|f\rangle\to|e\rangle$ and $|f\rangle\to|g\rangle$ contribute to the fluorescence
 from the double-excited state. 
Note that when the transition dipole  $\mu_{fg}$ vanishes, the fluorescence
solely comes from the transition $|f\rangle\to |e\rangle$ in $R^{(2)}$.

 To study the scaling with the pump intensity, we consider a case
 where both "signal" and "idler" modes have both
 vertical and horizontal polarizations with equal magnitudes and all molecular transition dipoles are aligned by $45^0$ angle to both directions so that
 ${\cal M}$ in Eq. 6 in the SI is independent of the field polarization.
 All terms in $R^{(2)}$ obtained after interchanging polarizations then make the same contribution 
 as the first term in $R^{(2)}$ and we can drop the polarization subscripts.
 The renormalized signal 
 $\tilde{R}(\omega_f)=\mbox{Re}~\{R^{(2)}(\omega_f)/{\cal M}^{fe_1;fe_1;fe;eg;e^\prime g;fe^\prime}\}$ is given by,
 \begin{widetext}
 \begin{subequations}
      \begin{equation}
  \label{eq-4}
  \hspace{-1cm}\tilde{R}(\omega_f) =  \mbox{Im}
\int\frac{d\omega_1d\omega_2d\omega_3}{(2\pi)^3} \frac{(48/\hbar^6)
D(\omega_1,\omega_2,\omega_3,\omega_1+\omega_2-\omega_3)}{|\omega_1+\omega_2-{\cal E}_{fg}+i\eta|^2(\omega_2-{\cal E}_{eg}+i\eta) 
(\omega_3-{\cal E}_{e^\prime g}-i\eta)(\omega_f-\omega_1-\omega_2+{\cal E}_{jg}-i\eta)}
  \end{equation} 
 where from Eq. (\ref{eq-1b})
 \begin{eqnarray}
 D(\omega_1,\omega_2,\omega_3,\omega_4)&=&\sum_{qq^\prime=i,s}
 \langle E^\dag_q(\omega_3)E_q(\omega_1)\rangle \langle E_{q^\prime}^\dag(\omega_4) E_{q^\prime}(\omega_2)\rangle + (\omega_1\Leftrightarrow\omega_2)\nonumber\\
 &+&\sum_{q\neq q^\prime=i,s}\sum_{q_1\neq q_1^\prime=i,s} \langle E_{q}(\omega_1)E_{q^\prime}(\omega_2)\rangle \langle E_{q}^\dag(\omega_3)E_{q_1^\prime}^\dag(\omega_4)\rangle.
 \end{eqnarray}
 \end{subequations}
  \end{widetext}
 For a weak pump, where only a single entangled-photon pair interacts with the molecule, only the last term in Eq, (7b) survives. 

\begin{figure}[h]
\includegraphics[scale=0.5]{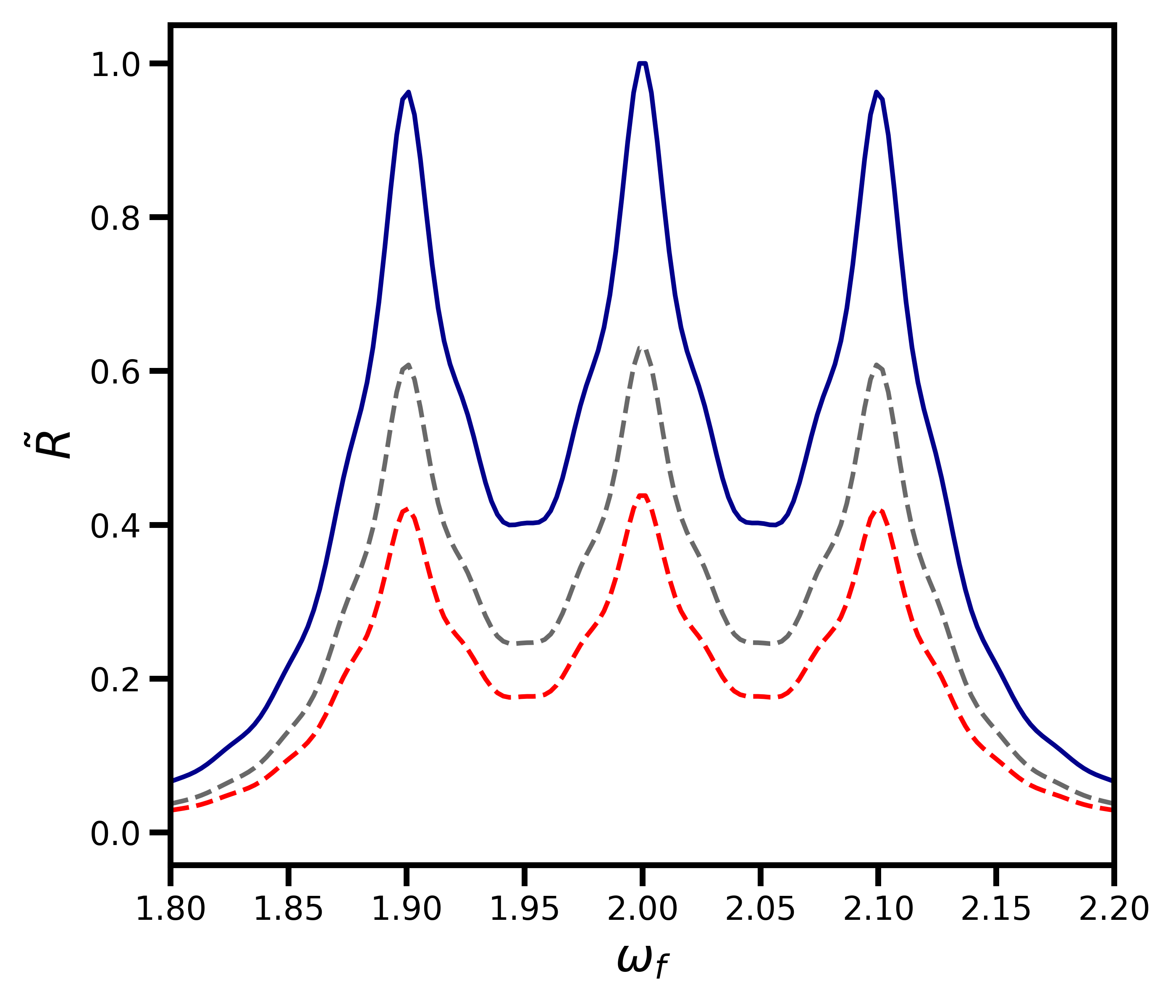}
\caption{ Normalized TPA signal from the model molecule (Eq. \ref{eq-4}).
The dashed-red (grey) curve is the inter-mode
contribution for $I_p$=3.73$\times$10$^{13}$ (9.33$\times$10$^{12}$)\,W/cm$^{2}$. 
Here $E_f=4 fs^{-1}, E_{e1}=1.9 fs^{-1},E_{e_2}= 2.0 fs^{-1}, E_{e_3}=2.1 fs^{-1}$, 
$\sigma_p=0.1 fs^{-1}$, $\eta$=0.015, $T_e$=13.33\,fs.} 
\label{fig-ws-vs-R2}
\end{figure}

 The first two terms, therefore, represent intra-mode contributions corresponding to the two rightmost diagrams in the middle panel in Fig. (\ref{fig-a}). 
 
 Figure (\ref{fig-ws-vs-R2}) depicts $\tilde{R}(\omega_f)$ for a model 
 molecule with one $|f\rangle$ state and three singly-excited states having energies
 $E_f=4 fs^{-1}, E_{e1}=1.9 fs^{-1},E_{e_2}= 2.0fs^{-1}, E_{e_3}= 2.1 fs^{-1}$ above
 the ground state $|g\rangle$. The three peaks represent transitions from 
 $|f\rangle$ to the three intermediate states. The inter-mode contribution is shown by the dashed curves for two different pump intensities. The relative weight of the inter-mode contribution depends on the pump amplitude and grows as the pump intensity is decreased. This is because the probability that the molecule interacts with two photons of an entangled-pair increases at lower intensities.

\begin{figure}[h]
\includegraphics[scale=0.6]{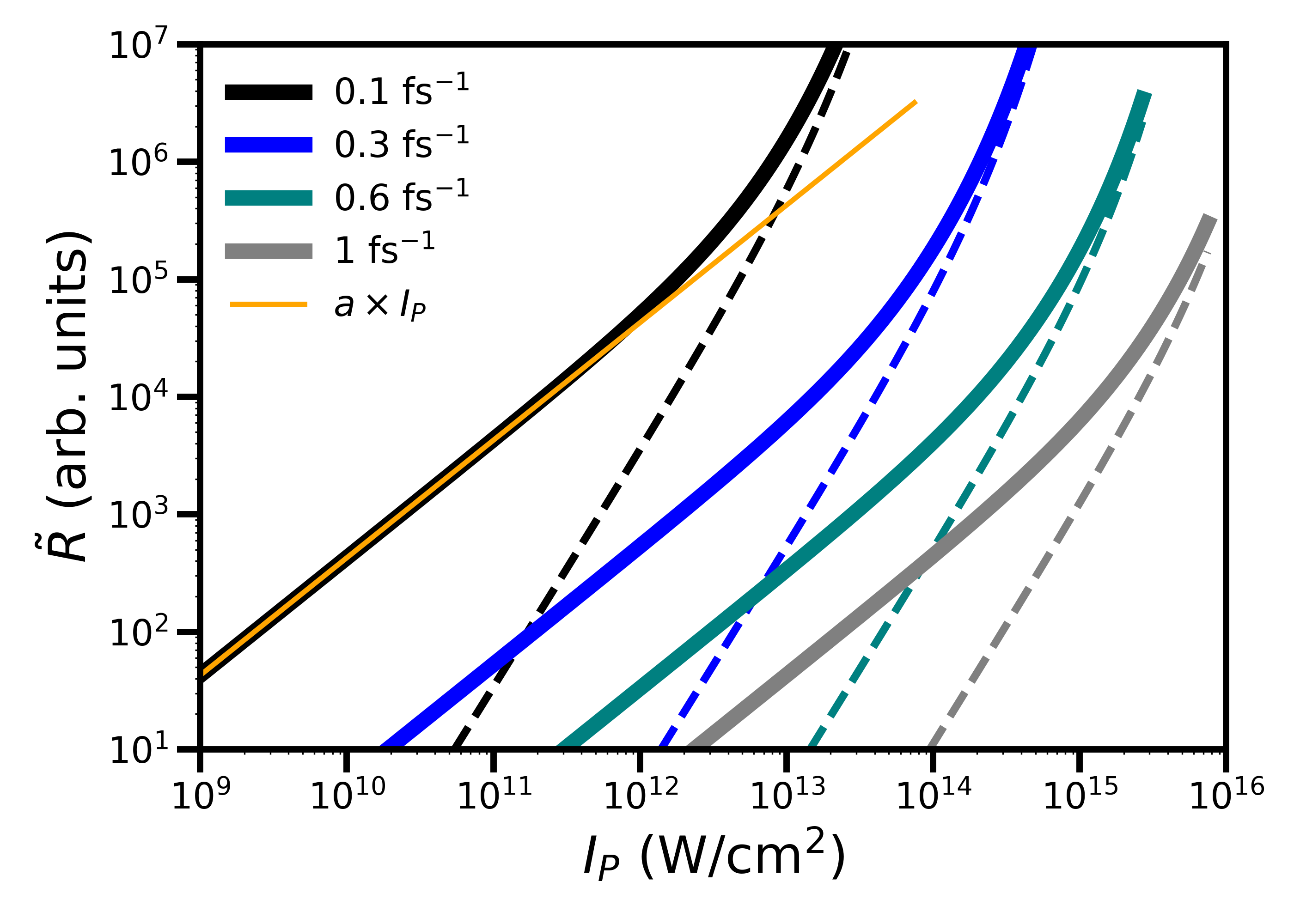}
\caption{Variation of the TPA signal for $\omega_f=2$ fs$^{-1}$ with pump-field intensity. The curves (top to bottom) denote the total signal calculated using Eq. (\ref{eq-4}) for pump band-widths $\sigma_p=0.1$\,fs$^{-1}$, $0.3$\,fs$^{-1}$,  $0.6$\,fs$^{-1}$, and $1.0$\,fs$^{-1}$, respectively. The dashed curves show intra-mode contributions. The orange line is the fit $\tilde{R}= 2.35\times 10^{-8} I_p$. The entanglement time is $T_e$=13.33\,fs, $\bar{n}_s=\bar{n}_i$=2.24,
$n_p$=2.361, and $\chi_{eff}^{(2)}$=4.6\,pm/V.}
\label{fig-EP-vs-R2}
\end{figure}

The variation of the TPA signal with the pump intensity $I_p = c\epsilon_0 n_p |\mathcal{E}_p|^2/2$, where  $\mathcal{E}_p$ is the pump pulse amplitude, and $n_p$ is the refractive index of the
PDC crystal for the pump-pulse, $c$ is the speed of light, and $\epsilon_0$ is the permittivity of free space, at the resonant 
frequency $\omega_s=2.0$ fs$^{-1}$is shown in Fig. (\ref{fig-EP-vs-R2}).
 %%%%%%%%
The inter-mode contributions grow linearly over a wide range
of pump intensity and dominates over the intra-mode
contributions that grow quadratically. However, for larger intensities, both
contributions grow nonlinearly, while the inter-mode contribution remains higher.
The TPA cross-section grows by orders of magnitude as the pump band-width is reduced,
however the range of pump intensity over which the inter-mode contribution dominates
decreases slowly with increasing band-width.
PDC parameters for the LiNbO$_3$ are considered in calculating the
TPA signal \cite{Shoji97josab}. The refractive indices for the considered
photon frequencies in the PDC crystal are obtained following the
procedure in Ref. \cite{edwards84oqe}. The value of the pump intensity beyond which nonlinear effects become significant for three commonly used PDC crystals at different pump spectral width are given in the Table.

\begin{table*}[ht!]
\centering
\begin{tabular}{|c|c|c|c|c|}
    \hline
    \diagbox{Crystal}{$\sigma_p$} & \textbf{0.1 (fs$^{-1}$)} & \textbf{0.3 (fs$^{-1}$)} & \textbf{0.6 (fs$^{-1}$)} & \textbf{1 (fs$^{-1}$)} \\ \hline
    \textbf{LiNbO$_3$ ($n_i$=2.24, $n_p$=2.36, $\chi^{(2)}_{eff}$=4.6\,pm/V)} & $5\times 10^{11}$  & $7.5\times 10^{12}$ & $5\times 10^{13}$ & $4\times 10^{14}$ \\ \hline
    \textbf{LiTaO$_3$ ($n_i$=2.14, $n_p$=2.23, $\chi^{(2)}_{eff}$=0.85\,pm/V)} & $ 1.5 \times 10^{13}$ & $2 \times 10^{14}$ & $1.5 \times 10^{15}$ & $8 \times 10^{15} $ \\ \hline
    \textbf{KNbO$_3$ ($n_i$=2.13, $n_p$=2.24, $\chi^{(2)}_{eff}$=10.8\,pm/V)} & $1.1 \times 10^{11}$ & $1.2 \times 10^{12}$ & $ 8 \times 10^{12} $ & $1.4 \times 10^{13}$ \\ \hline
\end{tabular}
\caption{Pump intensities (in W/cm$^2$) at which linear-to-nonlinear crossover appears for
different crystals and pulse widths are depicted.} \label{table}
\end{table*}
The inter-mode contributions reveal the quantum nature of light.
For example, information regarding the entanglement time and the 
linear variation of the signal over an order-of-magnitude larger 
intensity range. These contributions dominate at lower pump 
intensities but are masked by intra-mode processes at higher 
intensities, making it hard to take full advantage of the quantum 
light. In the next section, we propose a setup that removes the 
intra-mode contributions at all pump intensities. 
 
\subsection{Filtering out the intra-mode contributions}
We now present an experimental technique that filters out the unwanted (classical)
intra-mode contributions from the TPA signal. The inter- and intra-mode processes
have different contributions, as shown in Fig. (\ref{fig-ws-vs-R2}) and 
(\ref{fig-EP-vs-R2}), and the latter dominate at higher pump intensities.
Thus, in order to take 
full advantage of the quantum nature of the entangled light,
these contributions should be removed.
\begin{figure}
    \centering
    \includegraphics[scale=0.4]{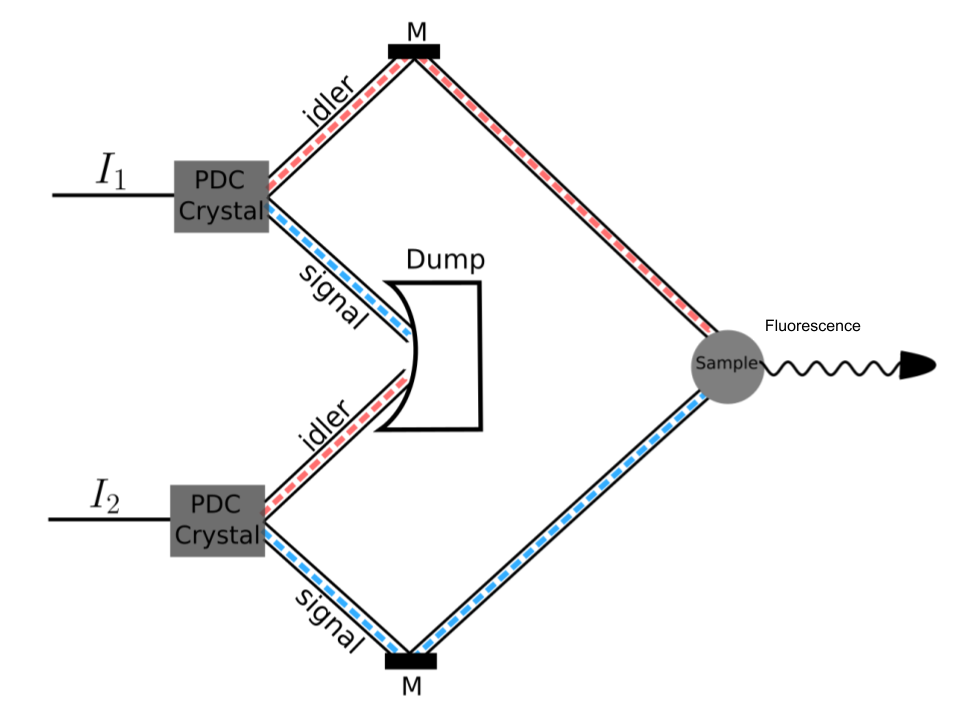}
    \caption{Schematic setup that can be used to
    generate TPA signal that only includes intra-mode
    contributions. The $s$ and $i$ photons from two different
    entangled photon pairs are utilized. The use of identical
    pump and the PDC crystal ensure the generated entangled
    photon pairs are statistically identical to those generated by a single PDC crystal but uncorrelated.}
    \label{fig:tpa_scheme}
\end{figure}

The inter-mode contributions vanish for a light field with no entangled modes.
We can thus replace the "signal" and "idler" pulses with two uncorrelated pulses
of the same intensity and then subtract this signal from that obtained using the 
entangled light. This will remove the intra-mode contributions.
However, two conditions must be satisfied: bandwidths and the statistics
(as determined by Glauber function) of the uncorrelated "signal" and "idler" 
modes must be the same as obtained from the actual PDC process.
The simplest way to achieve this would be by utilizing two identical PDC crystals to
generate statistically identical pairs of "signal" and "idler" pulses.
We then select the "signal" pulse from one pair and the "idler" pulse from the other pair.
These pulses are obviously not entangled, but are statistically identical to
the entangled pulses and, upon interaction with the molecule,
generate the signal, as depicted schematically in Fig. (\ref{fig:tpa_scheme}).
This signal will only have the intra-mode contributions since the
two modes are not correlated. The signal due to inter-mode contributions
is obtained by subtracting this signal from the total signal generated
using entangled light, i.e., the signal measured in the presence of only
one of the PDC crystals.      

\section{conclusion}
Two-photon-absorption with an entangled photon field is a weak 
process. The advantage of entangled light is limited to lower pump intensities where
contribution from interaction with entangled photons can outweigh
contribution from non-entangled photons by orders of magnitude.
This range of pump intensities can be enhanced by using a pump-pulse with
a broader band-width, at the cost of signal intensity. At higher intensities, both
contributions are almost equal but can be separated using the
proposed experimental scheme.

Finally, we note that (i) The effective Hamiltonian used in the present
study and elsewhere \cite{DayanPRA2007,Raymer2022}, 
can be derived in a 
clean way from an effective action of the low-frequency electromagnetic (EM) field for 
strong pump (the weak pump perturbative expansion of the effective action has been 
introduced in \cite{MatthiasPNAS2023}) by considering the complete
perturbative expansion for the 
solution of the Dyson equation and neglecting the fast-oscillating (non-RWA) terms,
whose contributions can be neglected, (ii) By applying the Kirchhoff-integral 
approach to the correlation functions of the field, together with the 
semiclassical (in linear optics known as eikonal) expressions for the Green 
functions of the Maxwell equations, one can explicitly express the entangled field
correlation functions at the experimental sample in terms of their at-the-PDC-crystal
counterpart, with those expressions carefully describing the optical experimental setup,
$e.g.$, lenses, mirrors, delay lines, etc., thus providing the actual values
(including pre-factors) of the measured signal. However, the aforementioned 
results go beyond the scope of this paper and will be addressed elsewhere.

\acknowledgments
We thank Victor M. Freixas for the helpful discussion. U.H. acknowledges support from the Science and Engineering Board, India, under Grant No. CRG/2020/0011100 and the Fulbright-Nehru Academic and Excellence Fellowship, 2023-24, sponsored by the U.S. Department of State and the United State-India Education Foundation. 
V.Y.C. and S.M. gratefully acknowledge the support of the US Department of Energy, Office of 
Science, Basic Energy Sciences Award DES0022134, which has primarily funded this work. 
S.M. gratefully acknowledges the support of the National Science Foundation through Grant No. CHE-2246379. 

\section*{Disclosure}
The authors declare no conflicts of interest.

\section*{Data Availability Statement}
Data underlying the results presented in this paper are not publicly available at this time but may be obtained from the authors upon reasonable request.

\appendix


\references

\bibitem{roadmap}
Shaul Mukamel, $et$ $al$, J. Phys. B: At. Mol. Opt. Phys. {\bf 53},  072002 (2020).

\bibitem{GillesPRA1993}
L. Gilles and P. L. Knight,  Phys. Rev. A {\bf 48}, 1582 (1993).

\bibitem{PRL1989}
J. Gea-Banacloche,  Phys. Rev. Lett. {\bf 62}, 1603 (1989).

\bibitem{OtEx2021}
T. Landes, M. G. Raymer, M. Allgaier, S. Merkouche, B. J.
Smith, and A. H. Marcus, Opt. Express {\bf 29}, 20022 (2021).

\bibitem{JimensasePRApp2021}
K. M. Parzuchowski, A. Mikhaylov, M. D. Mazurek, R. N.
Wilson, D. J. Lum, T. Gerrits, C. H. Camp, Jr., M. J. Stevens,
and R. Jimenez, Phys. Rev. Applied. {\bf 15}, 044012 (2021).

\bibitem{hickam22jpcl}
B.~P. Hickam, M.~He, N.~Harper, S.~Szoke, and S.~K. Cushing, \emph{J. Phys. Chem. Lett.}, vol.~13, no.~22, pp. 4934--4940, 2022.


\bibitem{corona22jpca}
S.~Corona-Aquino, O.~Calder{\'o}n-Losada, M.~Y. Li-G{\'o}mez, H.~Cruz-Ramirez, V.~{\'A}lvarez-Venicio, M.~D.~P. Carreon-Castro, R.~de J. Le{\'o}n-Montiel, and A.~B. U’Ren, \emph{J. Phys. Chem. A}, vol.~126, no.~14, pp. 2185--2195, 2022.

\bibitem{tabakaev21pra}
D.~Tabakaev, M.~Montagnese, G.~Haack, L.~Bonacina, J.-P. Wolf, H.~Zbinden, and R.~T. Thew, \emph{Phys. Rev. A}, vol.~103, no.~3, p. 033701, 2021.


\bibitem{LeeJPCB2006}
D.-Ik Lee, T. Goodson III, J. Phys. Chem. B, {\bf 110}, 25582 (2006).

\bibitem{GoodsonJPCL2013}
 L. Upton, M. Harpham, O. Suzer, M. Richter, S. Mukamel, and T. Goodson III,  J. Phys. Chem. Lett. {\bf 4}, 2046 (2013).
 
 \bibitem{GoodsonJPCL2017}
O. Varnavski, B. Pinsky, and T. Goodson III, J. Phys. Chem. Lett {\bf 8}, 388 (2017).

\bibitem{GoodsonJACS2009}
M. R. Harpham, \"{O}. S\"{u}zer, C. Ma, P. B\"{a}uerle, and T. Goodson III, J. Am. Chem. Soc {\bf 131}, 973 (2009).

\bibitem{GuzmanJACS2010}
A. R. Guzman, M. R. Harpham, \"{O}zg\"{u}n S\"{u}zer, M. M. Haley, and T. G. Goodson III, J. Am. Chem. Soc. {\bf 132}, 7840 (2010).

\bibitem{TabakaevPRL2022}
D. Tabakaev, A. Djorovic, L. La Volpe, G. Gaulier, S. Ghosh, L. Bonacina, J.-P. Wolf, H. Zbinden, and R.T. Thew, Phys. Rev. Lett. {\bf 129}, 183601 (2022).

\bibitem{LandesPRR2021}
T. Landes, M. Allgaier, S. Merkouche, B. J. Smith,
A. H. Marcus, and M. G. Raymer, Phys. Rev. Res. {\bf 3}, 033154 (2021). 

\bibitem{MatthiasPNAS2023}
M. Kizmann, H. K. Yadalam, V. Chernyak, and S. Mukamel, {\it Proc. Natl. Acad. Sci. U.S.A.}, {\bf 120}(30), e2304737120 (2023).

\bibitem{MonsalveJPCA2017}
J. P. V.-Monsalve, O. C.-Losada, M. N. Portela, and A. Valencia, 
J. Phys. Chem. A {\bf 121}(41), 7869 (2017).

\bibitem{BanaclochePRL1989}
J. Gea-Banacloche, Phys. Rev. Lett. {\bf 62}, 1603 (1989).

\bibitem{LandesArXive2024}
T. Landes, B. J. Smith, M. G. Raymer, arXiv:2404.16342


\bibitem{RaymerJCP2021}
M. G. Raymer, T. Landes, A. H. Marcus, J. Chem. Phys. {\bf 155}, 081501 (2021).

\bibitem{MikhaylovJPCL2022}
A. Mikhaylov, R. N. Wilson, K. M. Parzuchowski, M. D. Mazurek, C. H. Camp Jr., M. J. Stevens, and R. Jimenez, J. Phys. Chem. Lett. {\bf 13}, 1489 (2022)


\bibitem{DayanPRA2007}
B. Dayan, Phys. Rev. A {\bf 76}, 043813 (2007).

\bibitem{Raymer2022}
M. G. Raymer and T. Landes, Phys. Rev. A {\bf 106}, 013717 (2022).

\bibitem{svozilik18cp}
J.~Svozil{\'\i}k, J.~Pe{\v{r}}ina Jr., and R.~de J. Le{\'o}n-Montiel,  \emph{Chem. Phys.}, vol.~510, pp. 54--59, 2018.

\bibitem{UH-AvsQuantum2023}
U. Harbola, L. Candelori, J. R. Klien, V. Chernyak, and S. Mukamle, To appear in AVS Quantum (2023).



\bibitem{svozilik18josab}
J.~Svozil{\'\i}k, J.~Pe{\v{r}}ina, and R.~de J. Le{\'o}n-Montiel, \emph{JOSA B}, vol.~35, no.~2, pp. 460--467, 2018.


\bibitem{SilberhornNJP2013}
A. Christ, B. Brecht, W. Mauerer, and C. Silberhorn, New Journal of Physics {\bf 15}, 053038 (2013). 



\bibitem{NEGF08}
U. Harbola, S. Mukamel, Phys. Rep. {\bf 465} (5), 191 (2008).

\bibitem{NEGF06}
U. Harbola and S. Mukamel, J. Chem. Phys. {\bf 124}, 044106 (2006).

\bibitem{Bethe-Salpeter}
G. Onida, L. Reining, and A. Rubio,  Rev. Mod. Phys., {\bf 74}, 601 (2002).

\bibitem{Glauber}
R. J. Glauber,  Phys. Rev. {\bf 130}, 2529 (1963).

\bibitem{Shoji97josab}
I. Shoji, T. Kondo, A. Kitamoto, M. Shirane, and R. Ito, JOSA B {\bf 14}, 2268 (1997).
 
\bibitem{edwards84oqe}
G.~J. Edwards and M.~Lawrence, \emph{Opt Quant Electron}, vol.~16, pp. 373--375, 1984.
\end{document}

\title{Supplementary Information: Pump-intensity-scaling of Two-photon-Absorption and Photon Statistics of Entangled-Photon Fields
}

\author{Deependra Jadoun}
\affiliation{Department of Chemistry, University of California, Irvine, CA 92614, USA}
\affiliation{Department of Physics and Astronomy, University of California, Irvine, CA 92614, USA}
\author{Upendra Harbola}
\affiliation{Department of Inorganic and Physical Chemistry, Indian Institute of Science, Bangalore 560012,  India}
\author{Vladimir Y. Chernyak}
\affiliation{Department of Chemistry, Wayne State University, 5101 Cass Ave, Detroit, Michigan 48202, USA}
\affiliation{Department of Mathematics, Wayne State University, 656 W. Kirby, Detroit, Michigan 48202, USA}
\author{Shaul Mukamel}
\affiliation{Department of Chemistry, University of California, Irvine, CA 92614, USA}
\affiliation{Department of Physics and Astronomy, University of California, Irvine, CA 92614, USA}

\date{\today}

\maketitle

\vspace{-0.6cm}

\begin{widetext}
\section{Derivation of Eq. 1 in the main text}
\label {appendixA}

The molecular system interacting with the radiation field is represented by the Hamiltonian,
\begin{eqnarray}
\label{app-1}
H&=&\sum_{i}{\cal E}_iB_{ii} + \sum_{s}\hbar\omega_sE_s^\dag E_s+\sum_{s,i>j} (\mu_{ij}^s B_{ij}^\dag  E_s+h.c.) +{\cal H}
\end{eqnarray}
where ${\cal E}_i$ is the energy of the $i$th molecular state $i\rangle$, $B_{ij}=|j\rangle\langle j|$ is the exciton operator, $\mu_{ij}$ is the transition dipole-matrix element between states $|i\rangle$ and $|j\rangle$, $E_s^\dag (E_s)$ is the boson creation (annihilation) operator for the detected (fluorescence) field mode with frequency $\omega_s$. The third term represents 
 interaction of  the detected mode with the molecule, and the last term is the Hamiltonian of the incoming field that prepares the electronic 
 excitation and its interaction with the molecule. In our simulations, we assume an entangled-photon field described by Hamiltonian ${\cal H}$
 which needs not be specified at this point. 

The signal is defined by the rate of change of intensity in the detected mode $\omega_s$.
\begin{eqnarray}
\label{app-2}
R(\omega_s,t) &=& \frac{d}{dt}\langle E_s^\dag(t)E_s(t)\rangle=\frac{i}{\hbar}\langle [H,  E_s^\dag(t)E_s(t)]\rangle
= -\frac{2}{\hbar}\Im ~  \sum_{i>j} \mu_{ij}^s \langle B_{ij}^\dag(t) E_s(t)\rangle
\end{eqnarray}
where $\langle\cdot\rangle$ denotes average over the full  Hilbert space of  $H$. We assume weak interaction with the detected mode and compute 
the correlation $\langle B_{ij}^dag(t) E_s(t)\rangle$ to  leading order in the interaction. This gives,
\begin{eqnarray}
\label{app-3}
\langle B_{ij}^\dag(t) E_s(t)\rangle = \frac{-i}{\hbar}\sum_{i^\prime<j^\prime } \mu_{i^\prime j^\prime}^{s^\prime}
\int_{-\infty}^t d\tau
\langle B_{ij}^\dag(t) B_{i^\prime j^\prime}^\dag(\tau) \rangle
\langle E_s(t) E_{s^\prime}^\dag(\tau)\rangle,
\end{eqnarray}
where the time evolution is now in the interaction picture: molecular operators evolve with the Hamiltonian $\sum_{i}{\cal E}_iB_{ii} +{\cal H}$
and the field operators evolve with free-field Hamiltonian, $E_s(t)=E_s e^{-i\omega_s t}$, leading to 
$\langle E_s(t) E_{s^\prime}^\dag(\tau)\rangle=\delta_{ss^\prime} e^{-i\omega(t-\tau)}$. Substituting this in Eq. (\ref{app-3}), Eq. (\ref{app-2}) finally 
results in Eq. 1 of the main text.

\section{Contributions of the diagrams given in Fig. 1}
\label{appendixB}

For a fixed time ordering $\tau_1>\tau_2>\tau_3>\tau_4$, different dipole correlations can be evaluated in terms of molecular state energies as,
\begin{eqnarray}
\label{eq-1c}
&&\langle  B_{ijR}^\dag(t) B_{i^\prime j^\prime L}(\tau) B^\dag_{i_1j_1L}(\tau_1)B_{i_2j_2R}(\tau_2)\rangle =
\delta_{jj^\prime} \delta_{ii_2}\delta_{i^\prime i_1}\delta_{j_1g}\delta_{j_2g} e^{i{\cal E}_{ij}t}e^{-i{\cal E}_{i^\prime j^\prime}\tau} 
e^{-i{\cal E}_{i_2j_2}\tau_2} e^{i{\cal E}_{i_1j_1}\tau_1}\nonumber\\
 &&\langle T B_{ijR}^\dag(t) B_{i^\prime j^\prime L}(\tau) B^\dag_{i_1j_1L}(\tau_1)B^\dag_{i_2j_2L}(\tau_2)B_{i_3j_3R}(\tau_3)
B_{i_4j_4R}(\tau_4)\rangle= \delta_{jj^\prime}\delta_{i^\prime i_1}\delta_{ii_3}\delta_{j_3i_4}\delta_{j_4g}\delta_{i_2j_2}\delta_{j_2g}\nonumber\\
&&e^{-i{\cal E}_{i_4g}\tau_4} e^{-i{\cal E}_{i_3j_3}\tau_3} e^{i{\cal E}_{ij}t} e^{-i{\cal E}_{i^\prime j^\prime}\tau} e^{i{\cal E}_{i_1j_1}\tau_1} e^{i{\cal E}_{i_2g}\tau_2}
\nonumber\\
&&\langle  B_{ijR}^\dag(t) B_{i^\prime j^\prime L}(\tau) B^\dag_{i_1j_1L}(\tau_1)B_{i_2j_2L}(\tau_2)B_{i_3j_3L}^\dag(\tau_3)B_{i_4j_4R}(\tau_4)\rangle
=  \delta_{jj^\prime}\delta_{i^\prime i_1}\delta_{ii_4}\delta_{j_1j_2}\delta_{j_4g}\delta_{i_2i_3}\delta_{j_3g}\nonumber\\
&&e^{-i{\cal E}_{i_4g}\tau_4} e^{i{\cal E}_{ij}t} e^{-i{\cal E}_{i^\prime j^\prime}\tau} e^{i{\cal E}_{i_1j_1}\tau_1} e^{-i{\cal E}_{i_2j_2}\tau_2} 
e^{i{\cal E}_{i_3g}\tau_3}\nonumber\\
&&\langle TB_{ijR}^\dag(t) B_{i^\prime j^\prime L}(\tau) B^\dag_{i_1j_1L}(\tau_1)B_{i_2j_2L}(\tau_2)B_{i_3j_3L}^\dag(\tau_3)B_{i_4j_4R}(\tau_4)\rangle
=  \delta_{jj^\prime}\delta_{i^\prime i_1}\delta_{ii_4}\delta_{j_1j_2}\delta_{j_4g}\delta_{i_2i_3}\delta_{j_3g}\nonumber\\
&&e^{-i{\cal E}_{i_4g}\tau_4} e^{i{\cal E}_{ij}t} e^{-i{\cal E}_{i^\prime j^\prime}\tau} e^{i{\cal E}_{i_1j_1}\tau_1} e^{-i{\cal E}_{i_2j_2}\tau_2} e^{i{\cal E}_{i_3g}\tau_3}\nonumber\\
&&\langle  B_{ijR}^\dag(t) B_{i^\prime j^\prime L}(\tau) B^\dag_{i_1j_1L}(\tau_1)B_{i_2j_2R}(\tau_2)B_{i_3j_3R}^\dag(\tau_3)B_{i_4j_4R}(\tau_4)\rangle 
= \delta_{jj^\prime}\delta_{i_2 i}\delta_{i_1i^\prime}\delta_{j_2j_3}\delta_{i_4i_3}\delta_{j_4g}\delta_{j_1g}\nonumber\\
&&e^{-i{\cal E}_{i_4g}\tau_4} e^{i{\cal E}_{i_3j_3}\tau_3} e^{-i{\cal E}_{i_2 j_2}\tau_2} e^{i{\cal E}_{ij}t} e^{-i{\cal E}_{i^\prime j^\prime}\tau} e^{i{\cal E}_{i_1g}\tau_1}
\end{eqnarray}
where the subscripts  $i_1,i_2$, etc., denote molecular states, $|g\rangle, |e\rangle, |e^\prime\rangle,...,|f\rangle$ with energies 
${\cal E}_g, {\cal E}_e, {\cal E}_e^\prime,...,{\cal E}_f$, respectively.. 

By substituting Eq. (\ref{eq-1c}) in Eq. (3) in the main text and then in Eq. (1)
in the main text, the signal can be recast as, $R(\omega_s)=\sum_{n=1,6}\mbox{Re} ~R^{(n)}(\omega_s)$, with
\begin{equation}
    \begin{aligned}
        R^{(1)}(\omega_s) &= \frac{4}{\hbar^2}\sum_{\{\lambda\}} {\cal M}^{ij;ij;ig;ig}_{\lambda\lambda\lambda_1\lambda_2} 
\int\frac{d\omega_1}{2\pi} \frac{iD^{-+LR}_{\lambda_1\lambda_2}(\omega_1,\omega_1)}
{|\omega_1-{\cal E}_{ig}+i\eta|^2(\omega_s-\omega_1+{\cal E}_{jg}-i\eta)}\label{eq-2-1}
    \end{aligned}
\end{equation}
\begin{equation}
    \begin{aligned}
\hspace{-1cm}R^{(2)}(\omega_s) =& \frac{12}{\hbar^6}\sum_{\{\lambda\}} 
\int\frac{d\omega_1d\omega_2d\omega_3}{(2\pi)^3} \frac{(-i) {\cal M}^{fj;fj;fe;eg;e^\prime g;fe^\prime}_{\lambda\lambda\lambda_1\lambda_2\lambda_3\lambda_4}
D^{--++LLRR}_{\lambda_1\lambda_2\lambda_3\lambda_4}
(\omega_1,\omega_2,\omega_3,\omega_1+\omega_2-\omega_3)}{(\omega_2-{\cal E}_{eg}+i\eta) |\omega_1+\omega_2-{\cal E}_{fg}+i\eta|^2(\omega_3-{\cal E}_{e^\prime g}-i\eta)}\\ & \times \cfrac{1}{(\omega_s-\omega_1-\omega_2+{\cal E}_{jg}-i\eta)}
+(\lambda_1\Leftrightarrow \lambda_2) + (\lambda_3\Leftrightarrow \lambda_4) + (\lambda_1\Leftrightarrow \lambda_2, \lambda_3\Leftrightarrow \lambda_4 )
\label{eq-2-2}
    \end{aligned}
\end{equation}
\begin{equation}
    \begin{aligned}
\hspace{-1cm}R^{(3)}(\omega_s) =& \sum_{\{\lambda\}} 
\int\frac{d\omega_1d\omega_2d\omega_3}{(2\pi)^3} \frac{(24 i/\hbar^6) {\cal M}^{ij;ij;ij^\prime;i^\prime j^\prime;i^\prime g;ig}_{\lambda\lambda\lambda_1\lambda_2\lambda_3\lambda_4}
D^{--++LLLR}_{\lambda_1\lambda_2\lambda_3\lambda_4}
(\omega_1,\omega_1-\omega_2-\omega_3,\omega_2,\omega_3)}{(\omega_2+{\cal E}_{i^\prime g}-i\eta) (\omega_1-\omega_3+{\cal E}_{jg}-i\eta)(\omega_3-{\cal E}_{ij^\prime}
-{\cal E}_{ig}+i\eta)(\omega_3-{\cal E}_{ig}-i\eta) }\\ & \times \cfrac{1}{(\omega_3-\omega_s+{\cal E}_{j^\prime g}+i\eta)}
+(\lambda_1\Leftrightarrow \lambda_3) \label{eq-2-3}\\
    \end{aligned}
\end{equation}

\begin{equation}
    \begin{aligned}
       \hspace{-1cm} R^{(4)}(\omega_s) =& \sum_{\{\lambda\}} 
\int\frac{d\omega_1d\omega_2d\omega_3}{(2\pi)^3} \frac{(24 i/\hbar^6) {\cal M}^{e^\prime g;e^\prime g;eg;fe^\prime;fe;e^\prime g}_{\lambda\lambda\lambda_1\lambda_2\lambda_3\lambda_4}
D^{--++LLLR}_{\lambda_1\lambda_2\lambda_3\lambda_4}
(\omega_1,\omega_1-\omega_2-\omega_3,\omega_2,\omega_3)}{(\omega_2+{\cal E}_{fe}-i\eta) (\omega_1-\omega_3+{\cal E}_{e^\prime e}-i\eta)|\omega_3-{\cal E}_{e^\prime g} +i\eta|^2  (\omega_3-\omega_s+i\eta)}\\ &+(\lambda_1\Leftrightarrow \lambda_3) \label{eq-2-4}
    \end{aligned}
\end{equation}
\begin{equation}
    \begin{aligned}
        \hspace{-1cm}R^{(5)}(\omega_s) =& \sum_{\{\lambda\}} 
\int\frac{d\omega_1d\omega_2d\omega_3}{(2\pi)^3} \frac{(24 i/\hbar^6) {\cal M}^{ij;ij;ig;ij^\prime;i^\prime j^\prime;i^\prime g}_{\lambda\lambda\lambda_1\lambda_2\lambda_3\lambda_4}
D^{--++LRRR}_{\lambda_1\lambda_2\lambda_3\lambda_4}
(\omega_1,\omega_2,\omega_1+\omega_2-\omega_3,\omega_3)}{(\omega_3-{\cal E}_{i^\prime g}-i\eta) (\omega_1+\omega_2-{\cal E}_{j^\prime g}-i\eta)
|\omega_1-{\cal E}_{ig}+i\eta|^2 }\\ & \times \cfrac{1}{(\omega_1+\omega_s-{\cal E}_{ij}-{\cal E}_{i g}+i\eta)} +(\lambda_2\Leftrightarrow \lambda_4) \label{eq-2-5}
    \end{aligned}
\end{equation}
\begin{equation}
    \begin{aligned}
        \hspace{-1cm}R^{(6)}(\omega_s) =& \sum_{\{\lambda\}} 
\int\frac{d\omega_1d\omega_2d\omega_3}{(2\pi)^3} \frac{(24 i/\hbar^6) {\cal M}^{e^\prime g;e^\prime g;e^\prime g;fe;fe^\prime;e g}_{\lambda\lambda\lambda_1\lambda_2\lambda_3\lambda_4}
D^{--++LRRR}_{\lambda_1\lambda_2\lambda_3\lambda_4}
(\omega_1,\omega_2,\omega_1+\omega_2-\omega_3,\omega_3)}{(\omega_3-{\cal E}_{eg}-i\eta) (\omega_1-\omega_3-{\cal E}_{fg}-i\eta)
|\omega_1-{\cal E}_{e^\prime g}-i\eta|^2  (\omega_1-\omega_s+i\eta)}\\ &+(\lambda_2\Leftrightarrow \lambda_4)\label{eq-2-6}
    \end{aligned}
\end{equation}
where the sum over the repeated indices $i,j,i^\prime,j^\prime$ runs over all excited states of the molecule while sums over $e,e^\prime$ are restricted to 
the singly excited states, and $\{\lambda\}$ denotes a summation over all possible values of the polarizations for $\lambda,\lambda_i, i=1,2,3,4$. 
Note that  ${\cal M}^{ij;ij;ig;ig}_{\lambda\lambda\lambda_1\lambda_2} = (\mu_{ij}\cdot\epsilon_\lambda)(\mu_{ij}^*\cdot\epsilon_\lambda)(\mu_{ig}
\cdot\epsilon_{\lambda_1})(\mu_{ig}^*\cdot\epsilon_{\lambda_2})$ and 
${\cal M}^{fj;fj;fe;eg;e^\prime g;fe^\prime}_{\lambda\lambda\lambda_1\lambda_2\lambda_3\lambda_4}= 
(\mu_{fj}\cdot\epsilon_\lambda)(\mu_{fj}^*\cdot\epsilon_\lambda)
(\mu_{fe}\cdot\epsilon_{\lambda_1})(\mu_{eg}^*\cdot\epsilon_{\lambda_2}) (\mu_{e^\prime g}\cdot\epsilon_{\lambda_3}) 
(\mu_{fe^\prime}^*\cdot\epsilon_{\lambda_4})$, etc. The six terms in Eqs. (\ref{eq-2-1}) -(\ref{eq-2-6}) correspond to the six diagrams given in Fig. 1 in the main text. 

\section{Nonperturbative calculation of the entangled-field correlation functions}\label{appendixC}

Entangled ("signal-idler") photon pairs are created by the interaction of the pump field with a nonlinear PDC crystal.
The entangled-field output is fully characterized by various correlations of the field operators, known as one-particle propagators 
that are encoded in the effective action $S_{eff}$ of the field. The entangled-photon
state is determined by the amplitude $E$ of the pump-field and by the nonlinear (second-order $\chi^{(2)}$) response of the PDC crystal
of length $l$ along the $z$-axis. Extension of the crystal in the $x-$ and $y-$ directions is assumed to be infinite 
with respect to the wavelengths of the photons involved in the process. 
The effective action approach leads to the following effective Hamiltonian $H_{eff}$ for the entangled-field generation.
\begin{eqnarray}
\label{h-eff}
H_{eff}(z) &=&  \sum_{\alpha}\int\frac{d\omega}{2\pi}\left(\kappa_{i\alpha}(\omega)E^\dag_{i\alpha}(\omega)E_{i\alpha}(\omega)
+\kappa_{s\alpha}(\omega)E^\dag_{s\alpha}(\omega)E_{s\alpha}(\omega) \right)\nonumber\\
&+& \hbar\sum_{\alpha,\beta}\int\frac{d\omega d\omega^\prime}{(2\pi)^2} \chi^{(2)}(\omega,\omega^\prime) {\cal E}(\omega+\omega^\prime,z) 
E^\dag_{i\alpha}(\omega)E^\dag_{s\beta}(\omega^\prime) +h.c.
\end{eqnarray}
where $E^\dag_{i\alpha}(\omega)$ is the creation operator in the "idler" mode with polarization $\epsilon_\alpha, \alpha$ (horizontal "H" or vertical "V"), frequency 
$\omega$,  $\kappa_{i\alpha}(\omega)$ is the corresponding dispersion of the mode, ${\cal E}(\omega,z)$ is the amplitude of the pump field  
at the position $z$ along the propagation direction inside the crystal, and 
$\chi^{(2)}(\omega,\omega^\prime)\approx\frac{1}{2l}\sqrt{\frac{\omega_i\omega_s}{n_in_s}}\chi^{(2)}_{eff}(\omega,\omega^\prime)$, where $l$ is the length of the PDC crystal along the $z$-axis, $\omega_{i/s} (n_{i/s})$ is the central frequency (refractive index) of the idler/signal mode, and $\chi^{(2)}_{eff}$ is the 
is the second-order susceptibility of the PDC crystal. The polarization $\epsilon_\beta$ is perpendicular to $\epsilon_\alpha$,
$\epsilon_\alpha\cdot\epsilon_\beta=0$.
The second term in the above Hamiltonian, therefore, generates a Bell polarization state for the signal and the idler modes.

The evolution of modes inside the crystal is determined by the Heisenberg equations,
\begin{eqnarray}
\label{eom-1}
\frac{\partial}{\partial z} E_{i\alpha}(\omega,z) = -\frac{i}{\hbar} \kappa_{i\alpha}(\omega)  E_{i\alpha}(\omega,z) 
-\frac{i}{\hbar}\int \frac{d\omega^\prime}{2\pi} \chi^{(2)}(\omega,\omega^\prime) {\cal E}(\omega+\omega^\prime)  E_{s\beta}^\dag(\omega, z)
\end{eqnarray}
with the initial condition $E_{i\alpha}(\omega,-l)$ defined at $z=-l$.
$E_{s\alpha}$ is obtained by interchanging the indices $i$ and $s$ in Eq. (\ref{eom-1}).

The formal solution of Eq. (\ref{eom-1}) is 
\begin{eqnarray}
\label{formal-sol}
E_{i\alpha}(\omega,0) = U_{ii}^{\alpha\alpha}(\omega,\omega^\prime)E_{i\alpha}(\omega^\prime,-l)
+V_{is}^{\alpha\beta}(\omega,\omega^\prime)E_{s\beta}^\dag(\omega^\prime,-l)
\end{eqnarray}
where $\beta$ denotes a polarization $\epsilon_\beta$ perpendicular to $\epsilon_\alpha$. The entangled field propagators may be recast
in terms of the functions $U(\omega,\omega^\prime)$  and $V(\omega,\omega^\prime)$ introduced in Eq. (\ref{formal-sol}). For example, 
$D^{--LL}_{si;\alpha\alpha^\prime}(\omega,\omega^\prime)\equiv {\cal D}^{--}_{si;\alpha\alpha^\prime}(\omega,\omega^\prime)
=\langle E_{s\alpha}(\omega)E_{i\alpha^\prime}(\omega^\prime)\rangle$ is given by,
\begin{eqnarray}
\label{eq-2}
{\cal D}^{--}_{si;\alpha\alpha^\prime}(\omega,\omega^\prime) &=& \delta_{\alpha\beta}\int \frac{d\omega_1}{2\pi} U_{ss}^{\alpha\alpha}(\omega,\omega_1,0)
V_{is}^{\alpha^\prime\beta}(\omega^\prime,\omega_1,0), 
\end{eqnarray}
where $U_{ss}^{\alpha\alpha}(\omega,\omega^\prime,z)$ and $V_{is}^{\alpha\beta}(\omega,\omega^\prime,z)$ are the solutions of the coupled equations for 
$\tilde{U}_{ss}^{\alpha\alpha}(\omega,\omega^\prime,z)=e^{i\kappa_{s\alpha}(\omega)z}U_{ss}^{\alpha\alpha}(\omega,\omega^\prime,z)$ and 
\textit{}$\tilde{V}_{is}^{\alpha\beta}(\omega,\omega^\prime,z)=e^{i\kappa_{i\alpha}(\omega)z}V_{is}^{\alpha\beta}(\omega,\omega^\prime,z)$,
\begin{eqnarray}
\label{eom-UV1}
\frac{\partial}{\partial z}\tilde{U}_{ss}^{\alpha\alpha}(\omega,\omega^\prime,z) &=& 
-i\int\frac{d\omega_1}{2\pi} \chi^{(2)}(\omega,\omega_1){\cal E}(\omega+\omega_1)
e^{-i\Delta k_{si}^{\alpha\beta}(\omega,\omega_1)z}[\tilde{V}_{is}^{\beta\alpha}(\omega_1,\omega^\prime,z)]^*\nonumber\\
\frac{\partial}{\partial z}\tilde{V}_{is}^{\alpha\beta}(\omega,\omega^\prime,z) &=& 
-i\int\frac{d\omega_1}{2\pi} \chi^{(2)}(\omega,\omega_1){\cal E}(\omega+\omega_1)
e^{-i\Delta k_{is}^{\alpha\beta}(\omega,\omega_1)z}[\tilde{U}_{ss}^{\beta\beta}(\omega_1,\omega^\prime,z)]^*
\end{eqnarray}
with $\Delta k_{is}^{\alpha\beta}(\omega,\omega^\prime)=k_p(\omega+\omega^\prime)-\kappa_{i\alpha}(\omega)-\kappa_{s\beta}(\omega^\prime)$ 
and $k_p(\omega+\omega^\prime)$ being the momentum of the pump-pulse at frequency $\omega+\omega^\prime$. 
${\cal D}^{++}_{si\alpha\alpha^\prime}(\omega,\omega^\prime)$ is obtained from (\ref{eq-2}) by replacing the field annihilation operators by the 
corresponding creation operators. 

Note that  $U^{\alpha\alpha^\prime}$ is diagonal in the polarization indices $\alpha,\alpha^\prime$ while $V^{\alpha\beta}$ is off-diagonal. This,
together with Eq. (\ref{eq-2}), implies that the propagators ${\cal D}^{++}_{si\alpha\alpha^\prime}$ and ${\cal D}^{--}_{si\alpha\alpha^\prime}$ are non-zero
only for orthogonal polarizations $\epsilon_\alpha$ and $\epsilon_{\alpha^\prime}$. Similarly, ${\cal D}^{-+}_{ss\alpha\alpha^\prime}$
and ${\cal D}^{-+}_{ii\alpha\alpha^\prime}$ survive iff  the $\alpha$ and $\alpha^\prime$ polarizations are parallel. 
Finally, Eqs. (\ref{eom-UV1}) can be computed numerically for an arbitrary pump spectral envelop ${\cal E}(\omega)$, crystal response function
$\chi^{(2)}(\omega)$, and dispersions of the pump, idler and signal modes. 

Below we derive analytic expressions for a narrow  pump  ${\cal E}(\omega)=2\pi E_p\delta(\omega_p-\omega)$ with frequency $\omega_p$.
The main simplification in this case comes from the fact that the $z$-dependent coefficients in the differential equation (\ref{eom-UV1}) 
drops out and the evolution inside the crystal does not require $z$-ordering.  The $\omega_1$ integration in Eq. (\ref{eom-UV1}) can be performed trivially and we finally obtain, $U_{ss}^{\alpha\alpha}(\omega,\omega^\prime,0) = \delta(\omega-\omega^\prime){\cal U}_{ss}^{\alpha\alpha}(\omega,\bar{\omega})$ and $V_{is}^{\alpha\beta}(\omega,\omega^\prime,0) = \delta(\omega+\omega^\prime-\omega_p){\cal V}_{is}^{\alpha\beta}(\omega,\bar{\omega})$ where
$\bar{\omega}=\omega_p-\omega$ and,
\begin{eqnarray}
{\cal U}_{ss}^{\alpha\alpha}(\omega,\bar{\omega}) &=& e^{-i\frac{l}{2}\bar{\kappa}_{si}^{\alpha\beta}(\omega,\bar{\omega})}
\left[\mbox{cosh}\left(\frac{\kappa^{\alpha\beta}_{si} (\omega,\bar{\omega}) l}{2}\right)
+i\frac{\Delta k_{si}^{\alpha\beta}(\omega,\bar{\omega})}{\kappa^{\alpha\beta}_{si}(\omega,\bar{\omega})}
\mbox{sinh}\left(\frac{\kappa^{\alpha\beta}_{si}(\omega,\bar{\omega}) l}{2}\right)\right]\label{solution-Uii}\\
{\cal V}_{is}^{\alpha\beta}(\omega,\bar{\omega}) &=& -2i\frac{E_p\chi^{(2)}(\omega,\bar{\omega})}{\kappa_{is}^{\alpha\beta}(\omega,\bar{\omega})}
e^{\frac{i}{2}\bar{\kappa}_{si}^{\beta\alpha}(\bar{\omega},\omega)l}\mbox{sinh}\left(\frac{\kappa_{is}^{\alpha\beta}(\omega,\bar{\omega}) l}{2}\right)\label{solution-Vis}
\end{eqnarray}
where $\bar{\kappa}_{si}^{\alpha\beta}(\omega,\omega^\prime)=k_p(\omega_p)+\kappa_{s\alpha}(\omega)-\kappa_{i\beta}(\omega^\prime)$, and
$$\kappa_{si}^{\alpha\beta}(\omega,\omega^\prime)=
\sqrt{4|E_p|^2\chi^{(2)}(\omega,\omega^\prime)\chi^{(2)}(\omega^\prime,\omega)-(\Delta k_{si}^{\alpha\beta}(\omega,\omega^\prime))^2}.$$
$U_{ii}^{ss}(\omega,\omega^\prime,0)$ and $V_{si}^{\alpha\beta}(\omega,\omega^\prime,0)$ are obtained from Eqs. (\ref{solution-Uii}) and (\ref{solution-Vis}), respectively, by interchanging $i\Leftrightarrow s$. \\

{\bf Extension to a narrow pump pulse:} 
The above results obtained for a CW-pump can be extended approximately for a sufficiently narrow pump pulse by using the Wigner transformation of the pulse. Assuming that the pump amplitude varies slowly enough so that the phase, which varies with frequency $\omega_p$, is fast enough compared to the time-scale of the envelop (determined by the band-width), we extend the CW-pulse results as follows. First we re-write $U_{ss}^{\alpha\alpha}(\omega,\omega^\prime,0) = \int dt e^{i(\omega-\omega^\prime)t}{\cal U}_{ss}^{\alpha\alpha}(\omega,\omega_p-\omega^\prime)$ and $V_{is}^{\alpha\beta}(\omega,\omega^\prime,0) = \int dt e^{i(\omega+\omega^\prime-\omega_p)t} {\cal V}_{is}^{\alpha\beta}(\omega,\omega_p-\omega^\prime)$. This is exact for a CW-pump for any pump amplitude, $E_p$. We assume that the pump amplitude $E_p$ varies slowly enough so that $E_p$ can be replaced by its values at different times, which makes the functions ${\cal U}$ and ${|cal V}$ time dependent. Thus for a narrow but finite bandwidth pump we can use,
$$U_{ss}^{\alpha\alpha}(\omega,\omega^\prime,0) = \int dt~ e^{i(\omega-\omega^\prime)t}~{\cal U}_{ss}^{\alpha\alpha}(\omega,\omega_p-\omega^\prime,t)$$ $$V_{is}^{\alpha\beta}(\omega,\omega^\prime,0) = \int~ dt~ e^{i(\omega+\omega^\prime-\omega_p)t}~ {\cal V}_{is}^{\alpha\beta}(\omega,\omega_p-\omega^\prime,t).$$
These equations are used to evaluate the correlation functions for a narrow pump band-widths in the main text.

\subsection{The Two-Photon Correlation Function (TPCF) for a finite-band-width pump}

The TPCF $D_{si,HV}^{--}(\omega,\omega_1)$ in Eq. \ref{eq-2} is evaluated
numerically by solving Eqs. (\ref{eom-UV1}) using,
\begin{equation}
\begin{aligned}
        & \Delta k_{si}^{HV}(\omega,\omega_1) =  \cfrac{\omega-\Bar{\omega}_s}{v_{1,H}} + \cfrac{\omega_1-\Bar{\omega}_i}{v_{2,V}} \\
     & \Delta k_{is}^{VH}(\omega,\omega_1) =   \cfrac{\omega-\Bar{\omega}_i}{v_{2,V}} + \cfrac{\omega_1-\Bar{\omega}_s}{v_{1,H}}
\end{aligned}
\end{equation}
where $v_1$=$v_p-v_{a,X}$, and $v_2=v_p-v_{i,X}$ with $v_p$, $v_{s,X}$, and $v_{i,X}$
being group velocities of the pump,
the signal, and the idler photons, respectively, with polarization $\epsilon_X$, and 
$\Bar{\omega}_s$ ($\Bar{\omega}_i$) represents the center frequency of the signal (idler) photon. We take $l=20$\,$\mu$m crystal length, and the following spectral envelope of the pump field.
\begin{equation}
   {\cal E}(\omega) = E_p e^{-(\omega_p-\omega)^2/(2\sigma_p^2)}
\end{equation}
where $E_p$,  $\Bar{\omega_p}$, and $\sigma_p$ represent the
amplitude, central frequency,  and the spectral width of the field,
respectively.

\begin{figure}[h!]
\includegraphics[scale=0.6]{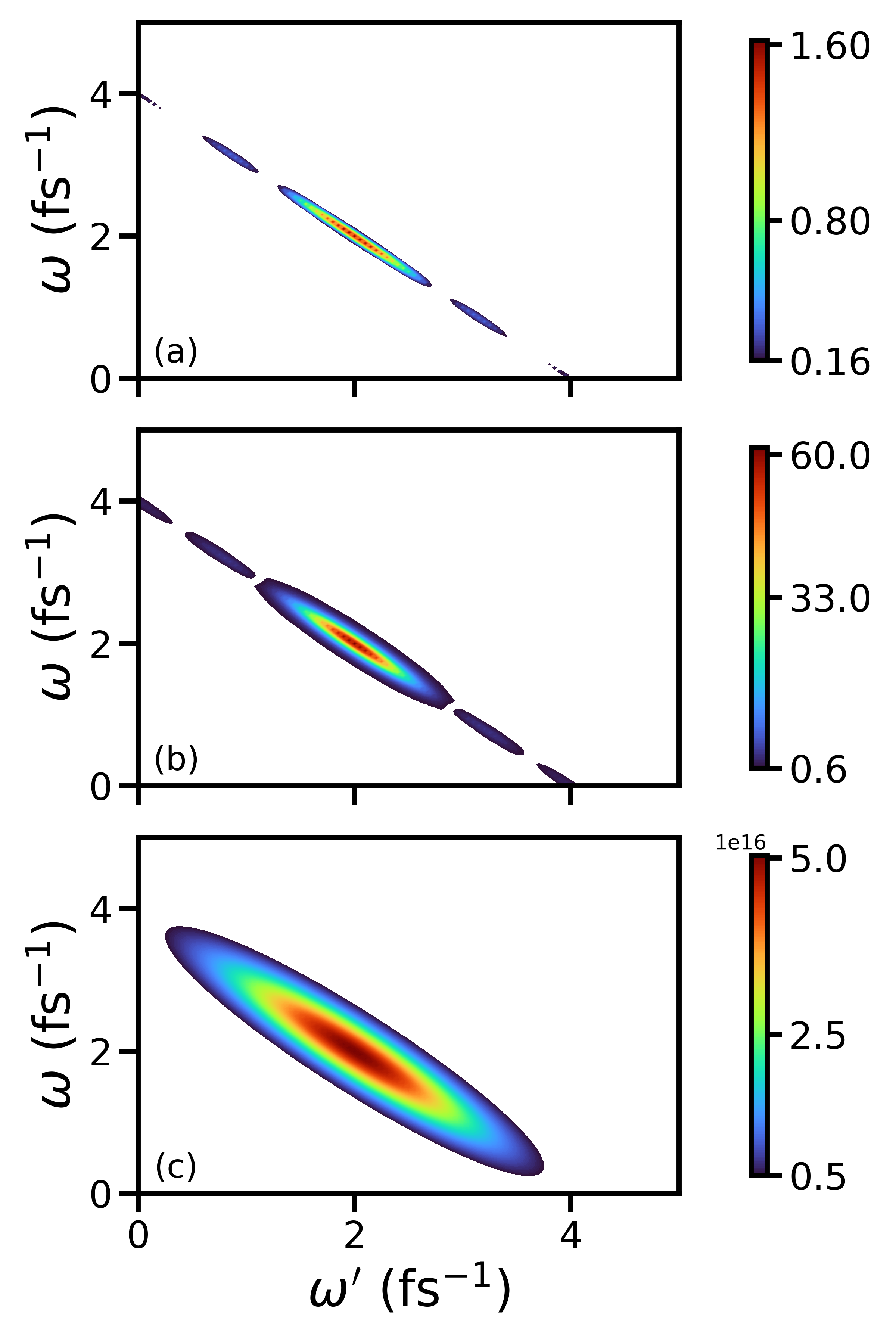}
\caption{
2D frequency-dispersed plots for $D_{si,HV}^{--}$ (Eq. 14) with $\sigma_p$=0.1\,fs$^{-1}$ and
(a) $E_p$=0.01\,a.u., (b) $E_p$=0.1\,a.u., and
(c) $E_p$=0.5\,a.u. are shown. Other system parameters: $v_{1,H}$=$v_{2,H}$=10\,$\mu$m$\cdot$fs$^{-1}$,
$v_{1,V}$=$v_{2,V}$=-10\,$\mu$m$\cdot$fs$^{-1}$ $\Bar{\omega_p}$=10\,fs$^{-1}$,
and $\Bar{\omega}_s$=5\,fs$^{-1}$.}
\label{fig:tpcf_ep}
\end{figure}
Figure \ref{fig:tpcf_ep} displays 2D frequency-dispersed plots for
$D_{si,HV}^{--}$ for a fixed pump bandwidth of $\sigma_p$=0.1\,fs$^{-1}$ and 
different pulse amplitudes ($E_p$).
Figure \ref{fig:tpcf_ep}(a) is constructed using
the weakest pump with $\mathcal{E}_p$=0.01\,a.u. 
Frequencies are clearly anti-correlated. For increased 
amplitude $E_p$=0.1\,a.u., the signal strength increases,
but the anti-correlation is still intact,
as shown in Fig. \ref{fig:tpcf_ep}(b).
Further increase to $E_p=1$\,a.u. leads to vanishing
of the features visible in the earlier two cases, as evident in
Fig. \ref{fig:tpcf_ep}(c). However, the frequency anti-correlation is still present in the signal. 
The amplitude of the correlation increases exponentially with the pump amplitude. 
\begin{figure}[h!]
\includegraphics[scale=.5]{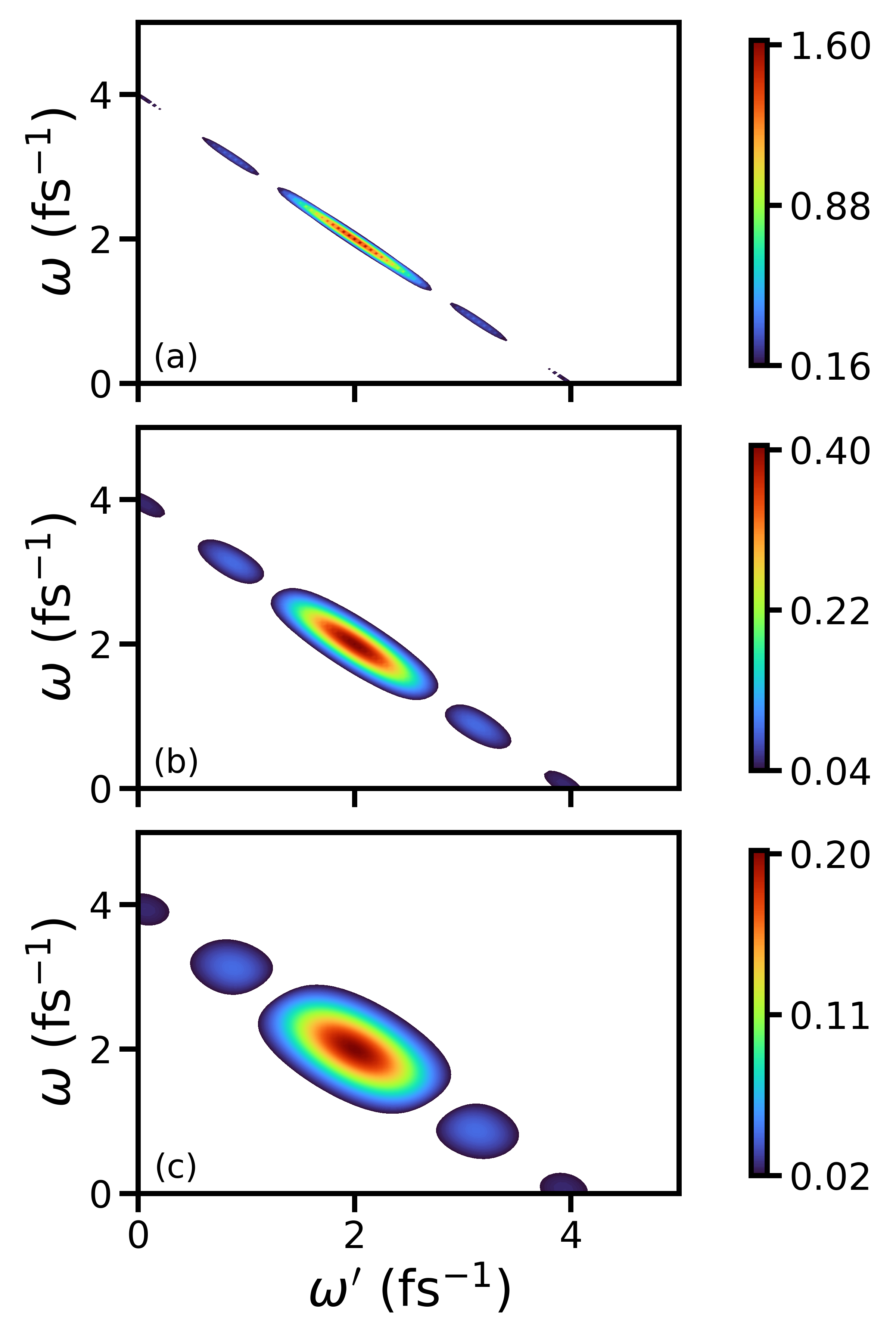}
\caption{ 2D frequency-dispersed plots for $D_{si,HV}^{--}$ (see Eq. 14) with $\mathcal{E}_p$=1\,a.u., and
(a) $\sigma_p$=0.1\,fs$^{-1}$, (b) $\sigma_p$=0.5\,fs$^{-1}$, and
(c) $\sigma_p$=1.0\,fs$^{-1}$ are shown. Other system parameters: $v_{1,H}$=$v_{2,H}$=10\,$\mu$m$\cdot$fs$^{-1}$,
$v_{1,V}$=$v_{2,V}$=-10\,$\mu$m$\cdot$fs$^{-1}$ $\Bar{\omega_p}$=10\,fs$^{-1}$,
and $\Bar{\omega}_s$=5\,fs$^{-1}$.}
\label{fig:tpcf_sp}
\end{figure}
To study the impact of the pump-width on the frequency anti-correlation, we
present in Fig. \ref{fig:tpcf_sp}  frequency dispersed plots of $D_{si,HV}^{--}$ for a fixed pump amplitude
but different band-widths. For a narrowband pump with $\sigma_p$=0.1\,fs$^{-1}$, the
frequency anti-correlation is still present, although limited
to the central blob, as shown in Fig. \ref{fig:tpcf_sp}(a).
However, for a larger pulse width with
$\sigma_p$=1\,fs$^{-1}$, the signal spreads into the entire frequency range with weak
features of anti-correlation, as shown in Fig. \ref{fig:tpcf_sp}(b).
For a broadband pulse with $\sigma_p$=1\,fs$^{-1}$, the frequency 
entanglement vanishes and the signal covers
the entire frequency range with
no signs of anti-correlation, as evident in Fig. \ref{fig:tpcf_sp}(c).
In addition, as can be seen by comparing the three panels in
Fig. \ref{fig:tpcf_sp}, the intensities increase significantly with the pulse width.

\subsection{$g^{(2)}(\tau)$ with zero bandwidth pump pulse}
\label{sec-highE}

For a zero-band-width pump, Eqs. (\ref{solution-Uii}) and (\ref{solution-Vis}) can be solved for the dynamical Eq. (\ref{eom-UV1}). $D_{si\alpha\alpha^\prime}^{--}(\omega,\omega^\prime)$ in Eq. (\ref{eq-2}) may be then expressed as (we remove polarization indices assuming that the "idler" and "signal" modes have fixed orthogonal polarizations),
\begin{eqnarray}
    \label{Dmm-app}
    D^{--}_{si}(\omega,\omega^\prime) &=& -\frac{2i}{h}\delta(\omega+\omega^\prime-\omega_p) \frac{\chi^{(2)}E_p}{\kappa_{is}(\omega\bar{\omega})}e^{-i\frac{l}{2}(\bar{\kappa}_{si}(\omega,\bar{\omega})-\bar{\kappa}_{si}(\bar{\omega},\omega))}
    \mbox{sinh}\left(\frac{l}{2}\kappa_{is}(\bar{\omega},\omega)\right)\nonumber\\
    &\times&\left(\mbox{cosh}\left(\frac{l}{2}\kappa_{si}(\omega,\bar{\omega})\right)+i\frac{\Delta k_{si}(\omega,\bar{\omega})}{\kappa_{si}(\omega,\bar{\omega})}\mbox{sinh}\left(\frac{l}{2}\kappa_{si}(\omega,\bar{\omega})\right)\right)
\end{eqnarray}
where $\bar{\omega}=\omega_p-\omega$ and the phase miss-match $\Delta k_{si}(\omega,\bar{\omega})$ is approximated using Taylor expansion $\kappa_{s\alpha}(\omega)\sim \kappa_{s\alpha}(\omega_s)+(\omega-\omega_s)T_{s\alpha}/l$ and $\kappa_{i\alpha}(\omega^\prime)\sim \kappa_{i\alpha}(\omega_i)+(\omega-\omega_i)T_{i\alpha}/l$ around the central frequencies $\omega_i$ and $\omega_s$ of the "idler" and "signal" photons, respectively, to obtain, $\Delta k_{si}(\omega,\bar{\omega})\approx -\frac{\Delta T}{l}(\omega-\omega_s)$ and $\Delta k_{is}(\omega,\bar{\omega})\approx \frac{\Delta T}{l}(\omega-\omega_i)$ with $\Delta T=T_s-T_i$ being the entanglement time. 
Similarly, for the intra-mode correlations, we get
\begin{eqnarray}
    \label{Dpm-app}
    D^{+-}_{ii}(\omega,\omega^\prime)&=& 4\delta(\omega-\omega^\prime) \left|\frac{E_p\chi^{(2)}}{\kappa_{si}(\omega,\bar{\omega})}\mbox{sinh}\left(\frac{\kappa_{si}(\omega,\bar{\omega})}{2}l\right)\right|^2. 
\end{eqnarray}

Note that for the narrow band-width pump, all intra-mode (inter-mode) propagators are diagonal (anti-diagonal) in the frequency space. This allows us to simplify the expression for $g^{(2)}(\tau)$ using Eq. (6) in the main text. 
\begin{eqnarray}
\label{g20-app}
    g^{(2)}(\tau)&=&1+\frac{1}{S(\tau)}\int \frac{d\omega d\omega^\prime}{(2\pi)^2}\sum_{q q^\prime=i,s}D^{+-}_{qq}(\omega,\omega) D^{+-}_{q^\prime q^\prime}(\omega^\prime,\omega^\prime) e^{-i(\omega-\omega^\prime)\tau}\nonumber\\
    &+&\sum_{q\neq q^\prime}\sum_{q_1\neq q_1^\prime}\frac{1}{S(\tau)}\int \frac{d\omega d\omega^\prime}{(2\pi)^2} D_{qq^\prime}^{++}(\omega,\omega)D_{q_1q_1^\prime}^{--}(\omega^\prime,\omega^\prime)e^{i(\omega_p-\omega-\omega^\prime)\tau}.
\end{eqnarray}

For simplicity, we assume a degenerate PDC process $\omega_i=\omega_s$. In this case, $g^{(2)}(\tau)$ simplifies to,
\begin{eqnarray}
    \label{g20a-app}
     g^{(2)}(\tau)&=& 1+ \frac{1}{S(\tau)} \left(\frac{8 (\chi E_p l)^2}{\Delta T}\right)^2\left| 
     \int\frac{d\omega}{2\pi} \frac{\mbox{sinh}^2(\frac{1}{2}\sqrt{(2\chi E_pl)^2-\omega^2})}{(2\chi E_pl)^2-\omega^2}e^{i\omega\frac{\tau}{\Delta T}}\right|^2  \nonumber\\
     &+&\frac{1}{S(\tau)}\left(\frac{2\chi E_p l}{\Delta T}\right)^2\left|\int\frac{d\omega}{2\pi} \frac{\mbox{sinh}(\sqrt{(2\chi E_pl)^2-\omega^2})}{\sqrt{(2\chi E_pl)^2-\omega^2}}e^{i\omega\frac{\tau}{\Delta T}}\right|^2\nonumber\\
     &\approx& 1+\frac{1}{\pi S(\tau)}\left(\frac{(\mbox{sinh}(\chi E_p l))^2}{\alpha_1\Delta T}\right)^2 e^{-\frac{1}{2}\left(\frac{\tau}{\alpha_1 \Delta T}\right)^2}
     + \frac{1}{2\pi S(\tau)}\left(\frac{\mbox{sinh}(2\chi E_pl)}{\alpha_2\Delta T}\right)^2 e^{-\frac{(\tau+\tau_0)^2+(\tau-\tau_0)^2}{(\alpha_2\Delta T)^2}}.
\end{eqnarray}
where in the second line we have used the approximation of steepest descent method to evaluate the integral, $\alpha_n=\sqrt{n\chi E_pl~coth(n\chi E_pl)-1}/(2\chi E_pl), n=1,2$, and $\tau_0=T_i+T_s$. The normalization $S(\tau)={\cal S}^2$ is independent of delay with ${\cal S}=2\int \frac{d\omega}{2\pi}|V_{si}(\omega,\omega)|^2$, which gives,
${\cal S}=\frac{(\mbox{sinh}(\chi E_p l))^2}{\sqrt{\pi}\alpha_1|\Delta T|}$. Substituting this in the above equation for $g^{(2)}(\tau)$, we get
\begin{eqnarray}
    \label{g20b-app}
    g^{(2)}(\tau)=1+ e^{-\frac{1}{2}\left(\frac{\tau}{\alpha_1\Delta T}\right)^2}+
    2\left(\frac{\alpha_1}{\alpha_2}\mbox{coth}(\chi E_p l) \right)^2~ e^{-\frac{(\tau+\tau_0)^2+(\tau-\tau_0)^2}{(\alpha_2\Delta T)^2}}.
\end{eqnarray}
The intra-mode contribution decays to unity while the inter-mode contribution decays to zero at large $\tau$. For $\tau=0$, the intra-mode part is independent on the pump intensity and is equal to $2$, while the inter-mode contribution decays rapidly with increasing pump intensity for small values of pump intensities, $E_p<1/(\chi l)$, and saturates to unity for large pump intensities.

    % \kappa_{si}^{\alpha\beta}(\omega,\omega^\prime)\sim 2E_p\chi \left[1-\left(\frac{\Delta T(\omega-\omega_s)}{E_p\chi l}\right)^2\right].
    % U_{ss}^{\alpha\alpha}(\omega,\omega^\prime) &\sim& \frac{1}{2}e^{E_p\chi l} e^{-\frac{1}{8}\frac{(\Delta T(\omega-\omega_s))^2}{E_p\chi l}} e^{i\frac{\Delta T(\omega-\omega_s)}{2E_p\chi l}} e^{-i l\kappa_{s\alpha}(\omega_s)} e^{-\frac{i}{2}(T_i+T_s)(\omega-\omega_s)} \nonumber\\
    % V_{is}^{\alpha\alpha}(\omega,\omega^\prime) &\sim& -\frac{i}{2}e^{E_p\chi l} e^{-\frac{1}{8}\frac{(\Delta T(\omega-\omega_s))^2}{E_p\chi l}} e^{i\frac{\Delta T(\omega-\omega_s)}{2E_p\chi l}} e^{-i l\kappa_{s\alpha}(\omega_s)} e^{-\frac{i}{2}(T_i+T_s)(\omega-\omega_s)}

\end{widetext}

\bibliography{report}   % bibliography data in report.bib
\bibliographystyle{achemso}